\documentclass[aps,showpacs,preprintnumbers,amsmath,amssymb,twocolumn,superscriptaddress,floatfix,nofootinbib,10pt]{revtex4-1}

\usepackage[linesnumbered,lined,boxed,commentsnumbered,ruled,vlined]{algorithm2e}
\usepackage{algpseudocode}
\usepackage{amsfonts}
\usepackage{amsmath}
\usepackage{amssymb}
\usepackage{bm,bbm}
\usepackage{booktabs}
\usepackage{caption}
\usepackage{centernot}
\usepackage{colortbl}
\usepackage{comment}
\usepackage{dcolumn}
\usepackage{epstopdf}
\usepackage{float}
\usepackage[bottom]{footmisc}
\usepackage{graphicx}
\usepackage{hyperref}
\usepackage{cleveref}
\usepackage{mathrsfs}
\usepackage{mathtools}
\usepackage{multirow}
\usepackage{physics}
\usepackage{ragged2e}
\usepackage{subfigure}
\usepackage{tabularx}
\usepackage{tikz}
\usepackage{verbatim}
\usepackage{xcolor}
\usepackage{xkcdcolors}
\DeclareCaptionJustification{justified}{\justifying}
\hypersetup{colorlinks=true, citecolor=orange, urlcolor=blue, linkcolor=magenta}

\definecolor{forest}{rgb}{0.0,0.27,0.13}
\definecolor{yellowcream}{rgb}{1.0,1.0,0.7}
\definecolor{verdeacqua}{rgb}{0.55,0.83,0.78}
\definecolor{lilla}{rgb}{0.75,0.73,0.85}

\newcommand{\beq}{\begin{equation}}
\newcommand{\eneq}{\end{equation}}

\begin{document}

\title{Statistically validated projection of bipartite signed networks}

\author{Anna Gallo}
\email{anna.gallo@imtlucca.it}
\affiliation{IMT School for Advanced Studies, Piazza San Francesco 19, 55100 Lucca (Italy)}
\affiliation{INdAM-GNAMPA Istituto Nazionale di Alta Matematica `Francesco Severi', P.le Aldo Moro 5, 00185 Rome (Italy)}
\author{Fabio Saracco}
\affiliation{Centro Studi e Ricerche `Enrico Fermi' (CREF), Via Panisperna 89A, 00184 Rome (Italy)}
\affiliation{IMT School for Advanced Studies, Piazza San Francesco 19, 55100 Lucca (Italy)}
\affiliation{Institute for Applied Computing `Mauro Picone' (IAC), National Research Council, Via Madonna del Piano 10, 50019 Sesto Fiorentino (FI) (Italy)}
\author{Tiziano Squartini}
\affiliation{IMT School for Advanced Studies, Piazza San Francesco 19, 55100 Lucca (Italy)}
\affiliation{INdAM-GNAMPA Istituto Nazionale di Alta Matematica `Francesco Severi', P.le Aldo Moro 5, 00185 Rome (Italy)}
\date{\today}

\begin{abstract}
Bipartite networks provide a major insight into the organisation of many real-world systems. One of the most relevant issues encountered when modelling a bipartite network is that of facing the information shortage concerning intra-layer linkages. In the present contribution, we propose an unsupervised algorithm to obtain statistically validated projections of bipartite signed networks, according to which any two nodes sharing a statistically significant number of concordant (discordant) relationships are connected by a positive (negative) edge. 
Our algorithm outputs a matrix of link-specific $p-$values, from which a validated projection can be obtained upon running a multiple-hypothesis testing procedure. After testing our method on synthetic configurations output by a fully controllable generative model, we apply it to several real-world configurations: in all cases, non-trivial mesoscopic structures, induced by relationships that cannot be traced back to the constraints defining the employed benchmarks, hence revealing genuine traces of self-organisation, are detected.
\end{abstract}


\maketitle

\section{Introduction}\label{sec:intro}

Network theory has emerged as a powerful framework to model different kinds of real-world systems, by representing their units as \textit{nodes} and the interactions between them as \textit{links}. Out of the many types of edges that have been considered so far, the \textit{signed} one, offering the possibility of modelling positive as well as negative interactions, has recently seen its popularity revived~\cite{antal2006social,leskovec2010signed,zaslavsky2012mathematical,tang2016survey}.

Most works on signed configurations have focused on monopartite graphs, i.e. configurations with a unique set of nodes each of which can interact with any other one. The interest towards one-mode signed networks is ascribable to the so-called \textit{balance theory} (BT), introduced by Heider in~\cite{heider1946attitudes} and later formalised by Cartwright and Harary employing signed graphs~\cite{cartwright1956structural}.

BT states that a signed graph is \textit{strongly balanced} (SB) if all cycles contain an even number of negative edges: from a mesoscopic perspective, this implies that the network can be split in two groups with positive intra-modular and negative inter-modular links; in~\cite{davis1967clustering}, Davis spoke of \textit{weakly balanced} (WB) graphs, allowing for more than two groups: taken together, the SB and WB concepts define what is called \textit{Traditional Balance Theory} (TBT)~\cite{gallo2024assessing} - a framework that has found applications in contexts as diverse as the biological, ecological, economic and social ones~\cite{harary2002signed,ou2015detecting,iorio2016efficient,saiz2017evidence,gallo2024testing,gallo2024assessing,gallo2025patterns,candellone2025community,talaga2023polarization,teixeira2017emergence}.

Since, however, real-world networks often deviate from it, Doreian and Mrvar have proposed a generalisation named \textit{Relaxed Balance Theory} (RBT)~\cite{doreian2009partitioning,gallo2024assessing} which allows for positive inter-modular and negative intra-modular links as well.\\

\paragraph*{The concept of balance in the bipartite context.} The interest towards two-mode signed networks, i.e. configurations with two sets of vertices where (signed) connections can be established only between pairs of nodes belonging to different sets, is, instead, much more recent. One of the earliest attempts at adapting the notion of structural balance to the bipartite framework has been carried out in~\cite{mrvar2009partitioning}, where the difference between the roles played by `subjects' and `objects' in Heider's formulation of the BT was highlighted. In~\cite{derr2018congressional,derr2019balance}, the authors focused on the balance of bipartite cycles by considering the statistical significance of the shortest ones, known as `butterfly motifs', `X-motifs' or `$2\times2$ bicliques'~\cite{saracco2015randomizing}. In~\cite{huang2021signed}, the authors carried out three different analyses: first, they tested the degree of balance of real-world bipartite signed networks by employing the definition of `butterfly motifs' provided in~\cite{derr2019balance}. Second, they inferred the sign of the missing intra-layer edges by connecting any two nodes belonging to the same set with a `plus one' (`minus one') if they established either two positive or two negative links (one positive and one negative link) with the same set of vertices on the opposite layer. Finally, they tested the degree of balance of each layer by comparing the empirical percentages of triangles and butterflies with the ones expected under a null model shuffling the links of the original bipartite network - i.e. implementing the microcanonical version of the Free-topology Signed Random Graph Model introduced in~\cite{gallo2024testing}: as a result, the patterns turned out to be more balanced than expected. In~\cite{banerjee2012partitioning}, bipartite signed networks were employed to model a group of individuals' opinions about certain topics: the authors addressed the problem of partitioning both types of entities in two groups while maximising the number of positive intra-modular links and minimising the number of negative inter-modular links.\\

\paragraph*{Projecting bipartite unsigned networks.} One of the issues of major interest encountered when modelling bipartite networks is that of inferring the presence of a relationship between nodes belonging to the same layer in case a direct measurement of such a relationship is unfeasible (as for data gathering about friendships on social platforms~\cite{neal2014backbone}). The simplest way of solving the problem is linking any two nodes belonging to the same layer as long as they share at least one neighbour: however, this often results in a very dense network whose topological structure is almost trivial. A different recipe prescribes retaining the information on the number of common neighbours, i.e. to project a bipartite network into a monopartite weighted network~\cite{neal2014backbone}: this prescription, however, causes the nodes with larger degree to have larger strengths, thus masking the genuine, statistical relevance of the induced connections. Moreover, such a prescription lets spurious clusters of nodes emerge (e.g. cliques induced by the presence of - even - a single node connected to all the other vertices on the opposite layer).

Algorithms for retaining only the significant weights have been proposed to face this problem~\cite{neal2014backbone}. Many of them are based on a thresholding procedure, a major drawback of which lies in the arbitrariness of the chosen threshold~\cite{latapy2008basic,watts1998collective,derudder2005cliquishness}. A different approach prescribes calculating the statistical significance of the projected weights according to a properly-defined null model~\cite{serrano2009extracting}: the latter, however, encodes relatively little information about the original structure, thus being more suited to analyse natively monopartite networks. A similar-in-spirit approach identifies the backbone of a monopartite weighted projection with its Minimum Spanning Tree and its communities with the trees constituting the Minimum Spanning Forest~\cite{zaccaria2014taxonomy,caldarelli2012network}: the lack of a comparison with a benchmark, however, makes assessing the statistical relevance of the outcome difficult.

All the aforementioned approaches validate a projection \textit{a posteriori}. A different class of methods focuses on recipes to obtain statistically validated projections by estimating the tendency of any two nodes belonging to the same layer to share a certain number of neighbours: all such approaches define a similarity measure that either ranges between $0$ and $1$~\cite{linden2003amazon,bonacich1972factoring} or follows a probability distribution allowing for a $p-$value to be computed~\cite{mantegna2011statistically,gualdi2016statistically,saracco2017inferring,dianati2016maximum,neal2024strong}. While, in the first case, the application of an arbitrary threshold is still unavoidable, in the second case, prescriptions rooted in traditional statistics can be applied.

The approaches discussed so far lead to unsigned projections of bipartite unsigned networks. In~\cite{neal2014backbone,neal2020sign}, instead, the author carried out a two-sided test of hypothesis to decide whether any two members of the $108$th U.S. Senate co-sponsored either enough bills for a political alliance (positive link) to be inferred or not enough bills for political antagonism (negative link) to be inferred. The employed null model was named Stochastic Degree Sequence Model, a principled derivation of which was provided in~\cite{saracco2015randomizing,saracco2017inferring} where the benchmark was re-named Bipartite Configuration Model.

For a comprehensive review of methods to carry out pattern detection in bipartite networks, see~\cite{neal2024pattern}.\\

\paragraph*{Projecting bipartite signed networks.} The issue of projecting bipartite signed networks has been addressed to a much less extent. A first example is provided by~\cite{li2018modeling}, where the authors projected a user-item network onto the layer of users: the similarity of any two of them was quantified by calculating the scalar product of the corresponding rows of the biadjacency matrix~\cite{fouss2007random} and its statistical significance evaluated by comparing the empirical value with the one expected under a null model defined by randomly swapping two edges having the same sign. Under such a benchmark - which is the microcanonical version of the Free-topology Signed Configuration Model introduced in~\cite{gallo2024testing} - the authors found that both the MovieLens and the Netflix projections were highly balanced - the degree of balance being proxied by the number of triangles having an even number of negative edges. A second example is provided by~\cite{baltakys2023inference}, where the authors obtained signed projections by carrying out a statistical validation of the number of neighbours shared by any two nodes belonging to the same layer via a hypergeometric-binomial mixture distribution.\\

Hereby, we build upon previous contributions by extending the Exponential Random Graphs framework to include null models suitable for analysing binary undirected bipartite signed networks and employ them to obtain statistically validated signed projections. To address such a problem, we extend the algorithm proposed in~\cite{saracco2017inferring}, based upon the idea that any two nodes sharing a significantly large number of neighbours should be linked in the corresponding monopartite projection. More precisely, we propose two variants of it, according to the way ambivalent patterns (i.e. patterns constituted by two nodes and an item, liked by one node and disliked by the other) and missing ties (characterising patterns constituted by two nodes and an item, with just one node expressing an opinion about it) are treated.\\

The rest of the paper is organised as follows. First, we introduce a quantity to measure the similarity of any two nodes belonging to the same layer. Second, we derive its probability distribution according to each considered benchmark. Third, we consider each pair of nodes and quantify the statistical significance of their similarity. Fourth, we link only the ones surviving a multiple-hypothesis testing procedure. Fifth, we employ our methods to obtain signed projections of several different data sets. Finally, we comment on our results.

\begin{figure*}[t!]
\centering
\includegraphics[width=0.95\textwidth]{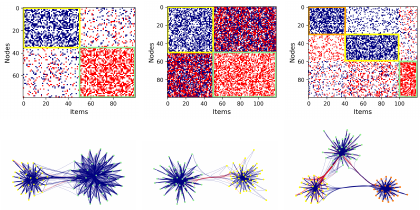}
\caption{Graphical representation of three synthetic configurations, generated by considering the values of the parameters reported in the main text. What we obtain confirms that \textit{i)} any two nodes within the same bipartite group share a significantly large number of concordant motifs that, in turn, induces a positive connection in the projection; \textit{ii)} any two nodes belonging to different bipartite groups may share either a significantly large number of discordant motifs, inducing a negative connection in the projection, or a significantly large number of concordant motifs, inducing a positive connection in the projection.}
\label{fig:4}
\end{figure*}

\section{Results}\label{sec:Results}

With the present contribution, we propose an unsupervised algorithm to obtain statistically validated projections of  binary undirected bipartite signed networks, according to which any two nodes sharing a statistically significant number of concordant (discordant) relationships are connected by a positive (negative) edge. Before applying it to a number of real-world configurations, let us, first, illustrate how it works on a toy model.

\begin{table*}[t!]
\centering
\begin{tabular}{l|c|c|c|c|c|c|c|c|c}  
\hline
\multicolumn{1}{l|}{} & \multicolumn{3}{c|}{FilmTrust} & \multicolumn{3}{c|}{U.S. Senate} & \multicolumn{3}{c}{U.S. House of Representatives} \\ 
\hline
\hline
\multicolumn{1}{c|}{} & $L^-$ & $L^+$ & $c$ & $L^-$ & $L^+$ & $c$ & $L^-$ & $L^+$ & $c$ \\
\hline
\hline
Zero-deflated projection - nai\"ve & 137.601 & 725.983 & 0.76 & 1.379 & 2.329 & 0.36 & \:\: 18.512 \:\: & \:\: 27.922 \:\: & 0.35 \\
\hline
Zero-deflated projection - global filter & 27.940 & 343.181 & 0.33 & 689 & 1.598 & 0.22 & 11.681 & 21.419 & 0.25 \\
\hline
Zero-deflated projection - local filter & 116.583 & 353.224 & 0.41 & 1.220 & 1.312 & 0.24 & 16.323 & 16.897 & 0.25 \\
\hline
\hline
Zero-deflated projection - nai\"ve & 0 & 1.135.278 & 1 & 34 & 10.406 & 1 & 14.379 & 117.976 & 1 \\ 
\hline
Zero-inflated projection - global filter & 77.024 & 667.448 & 0.65 & 3.942 & 5.041 & 0.86 & 47.839 & 67.631 & 0.87 \\
\hline
Zero-inflated projection - local filter & 56.803 & 62.057 & 0.10 & 2.897 & 2.337 & 0.50 & 40.982 & 32.693 & 0.56 \\
\hline
\end{tabular}
\caption{\label{table:1} From a purely empirical perspective, the datasets considered in the present contribution are characterised by a small link density and different percentages of positive and negative links: the connectance of \textit{U.S. Senate} and \textit{U.S. House of Representatives} amounts to $\simeq 0.1$ while $\simeq 55\%$ of links is positive; the connectance of \textit{FilmTrust} amounts to $\simeq 0.01$ while $\simeq 80\%$ of links is positive. Projections, instead, are characterised by a link density that depends on the scheme employed for filtering: overall, however, it is smaller than that of the na\"ive projections, the local filter cutting more edges than the global one.}
\end{table*}

\subsection{Projection of synthetic configurations}

In order to illustrate how our algorithm for projecting bipartite signed networks works, let us construct a generative model as follows. Let us consider the Bipartite Signed Stochastic Block Model (BiSSBM), induced by the finite scheme

\begin{equation}
b_{i\alpha}\sim
\begin{pmatrix}
-1 & 0 & +1\\
p_{g_ig_\alpha}^- & p_{g_ig_\alpha}^0 & p_{g_ig_\alpha}^+
\end{pmatrix},\quad i\in g_i,\:\alpha\in g_\alpha
\end{equation}
with $i=1\dots N$ and $\alpha=1\dots M$. Fig.~\ref{fig:4} provides a graphical representation of three configurations generated via the BiSSBM by considering parameters reading (left panels) $N=80$, $M=100$,
\begin{align}
    p^+=
\begin{pmatrix}
    0.5 & 0.08 \\
    0.08 & 0
\end{pmatrix}
\quad\text{and}\quad
p^-=
\begin{pmatrix}
    0 & 0.08 \\
    0.08 & 0.5
\end{pmatrix};    
\end{align}
(middle panels) $N=100$, $M=120$,
\begin{align}
    p^+=
\begin{pmatrix}
    0.5 & 0.4 \\
    0.4 & 0
\end{pmatrix}
\quad\text{and}\quad
p^-=
\begin{pmatrix}
    0 & 0.4 \\
    0.4 & 0.5
\end{pmatrix};
\end{align}
(right panels) $N=100$, $M=120$, 
\begin{align}
    p^+=
\begin{pmatrix}
    0.6 & 0.2 & 0.1\\
    0 & 0.6 & 0.1 \\
    0.15 & 0.1 & 0
\end{pmatrix}
\:\text{and}\:
p^-=
\begin{pmatrix}
    0 & 0 & 0.1\\
    0.2 & 0 & 0.1 \\
    0.1 & 0.2 & 0.6
\end{pmatrix}.
\end{align}

In the first case (left panels), the nodes belonging to $g_i=1$ ($g_i=2$) like the items belonging to $g_\alpha=1$ ($g_\alpha=2$) but establish few connections with the items belonging to $g_\alpha=2$ ($g_\alpha=1$); in the second case (middle panels), the nodes belonging to $g_i=1$ ($g_i=2$) like the items belonging to $g_\alpha=1$ ($g_\alpha=2$) but the density of inter-group connections is much larger; in the third case (right panels), the nodes belonging to the first (second) group like the items belonging to the first (second) group, the nodes belonging to the third group dislike the items belonging to the third group and the density of inter-group connections resembles the one characterising the first case. Irrespectively from the exact values of the parameters defining them, however, we expect the first two synthetic configurations to induce projections obeying the statistical variant of the traditional balance theory and the third configuration to induce a projection obeying the statistical variant of the relaxed balance theory~\cite{gallo2024assessing,doreian2009partitioning}.

Let us, now, project our synthetic configurations by employing the BiSRGM-FT, i.e. the global statistical benchmark induced by the zero-deflated scheme. Following~\cite{gallo2024assessing}, one can adopt an `agnostic' attitude and explore the mesoscale organisation of a projection without aligning with any specific conceptual framework: a principled approach to achieve such a goal is that of minimising the \textit{Bayesian Information Criterion} (BIC) reading

\begin{align}
\text{BIC}&=\kappa\ln V-2\ln\mathcal{L};
\end{align}
the first addendum proxies the \textit{complexity} of a model with the number of its parameters, $\kappa$, the second addendum proxies the \textit{accuracy} of a model with its log-likelihood, $\ln\mathcal{L}$, and $V=N(N-1)/2$ accounts for the system dimensions. Since we aim at describing a projection mesoscale organisation, a natural choice is that of adopting the Signed Stochastic Block Model (SSBM), defined by the likelihood function

\begin{align}
\mathcal{L}_\text{SSBM}=&\prod_{r=1}^k(p_{rr}^+)^{L_{rr}^+}(p_{rr}^-)^{L_{rr}^-}(1-p_{rr}^+-p_{rr}^-)^{\binom{N_r}{2}-L_{rr}}\nonumber\\
&\prod_{r=1}^k\prod_{\substack{s=1\\s>r}}^k(p_{rs}^+)^{L^+_{rs}}(p_{rs}^-)^{L^-_{rs}}(1-p_{rs}^+-p_{rs}^-)^{N_rN_s-L_{rs}}
\end{align}
and a number of parameters $\kappa_\text{SSBM}=k(k+1)$, amounting at twice the number of modules, $k$, into which the projection can be partitioned. Naturally, $N_r$ is the number of nodes constituting block $r$, $L_{rr}^+=p_{rr}^+N_r(N_r-1)/2$ is the number of positive links within block $r$, $L_{rr}^-=p_{rr}^-N_r(N_r-1)/2$ is the number of negative links within block $r$, $L_{rs}^+=p_{rs}^+N_rN_s$ is the number of positive links between blocks $r$ and $s$ and $L_{rs}^-=p_{rs}^-N_rN_s$ is the number of negative links between blocks $r$ and $s$.

\begin{figure*}[t!]
\centering
\includegraphics[width=\textwidth]{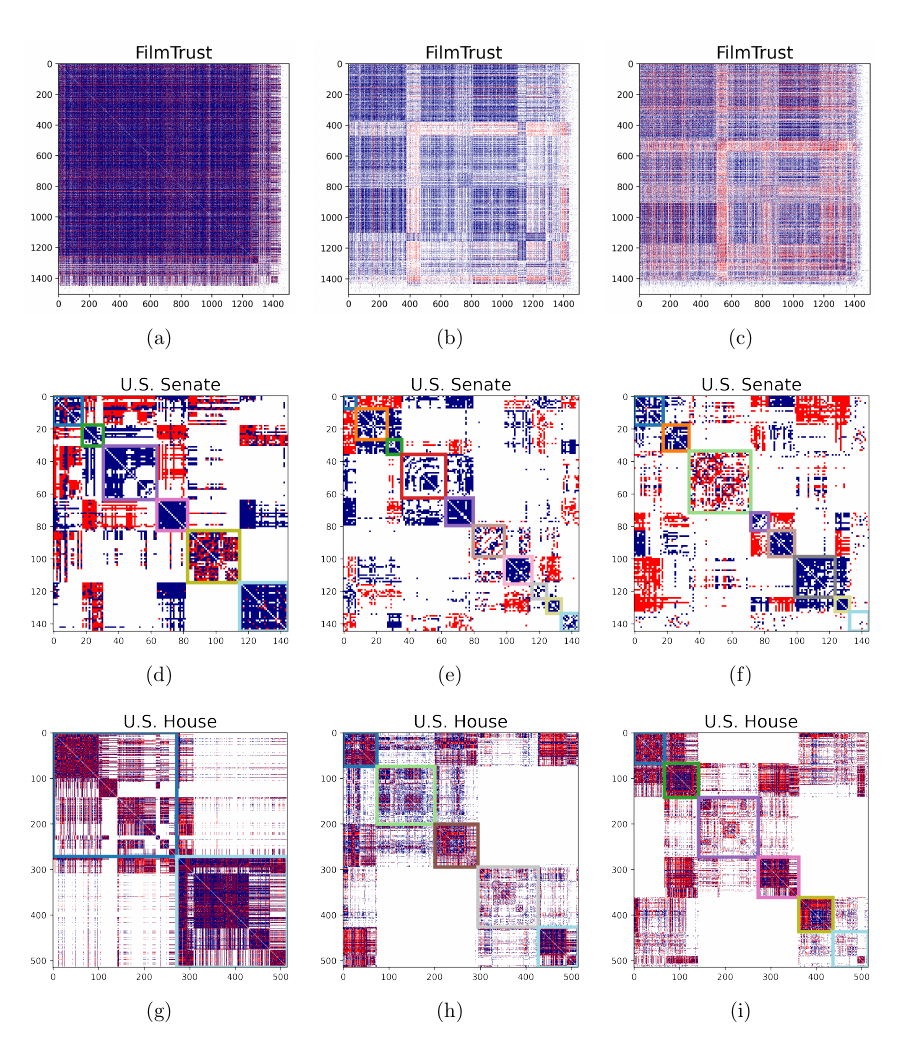}
\caption{Pictorial representation of the adjacency matrices of the projections of FilmTrust (panels (a)$-$(c)), U.S. Senate (panels (d)$-$(f)) and U.S. House of Representatives (panels (g)$-$(i)), obtained within the zero-deflated scheme, i.e. the na\"ive ones (panels (a), (d), (g)), the ones induced by the BiSRGM-FT (panels (b), (e), (h)) and the ones induced by the BiSCM-FT (panels (c), (f), (i)). Entries equal to $-1$ are coloured in red, entries equal to $0$ are coloured in white, entries equal to $+1$ are coloured in blue. The rows and columns of these adjacency matrices are re-ordered on the basis of the mesoscopic structures spotted by minimising BIC.}
\label{fig:5}
\end{figure*}

What we obtain confirms what we expect, i.e. that any two nodes within the same bipartite group share a significantly large number of concordant motifs that, in turn, induces a positive connection in the projection. On the other hand, any two nodes belonging to different bipartite groups may share either a significantly large number of discordant motifs that, in turn, induces a negative connection in the projection or a significantly large number of concordant motifs that, in turn, induces a positive connection in the projection.

\subsection{Projection of real-world configurations}

Let us, now, consider some real-world networks.\\

\paragraph*{U.S. Senate and U.S. House of Representatives.} The first two datasets that we consider, described in~\cite{derr2019balance} and further analysed in~\cite{derr2018congressional,huang2021signed}, are the output of the \textit{GovTrack.us} project~\cite{govtrackdataset} and collect vote records from the $1$st to the $10$th Congress of the United States. The nodes on the first layer are either senators or representatives while the nodes on the second layer are bills: a positive (negative) link between a senator/representative and a bill indicates that the senator/representative has voted `Yes' (`Nay') for that bill.\\

\paragraph*{FilmTrust.} The third dataset that we consider is the output of the \textit{FilmTrust} project~\cite{filmtrust,konectfilmtrust} and collects rating data from an online community whose users assign a score, i.e. 1, 2, 3, 4, to a number of movies. The nodes on the first layer are users while the nodes on the second layer are movies. We obtain a binary signed version of this dataset by assigning a $-1$ to all the edges whose weight is either 1 or 2 and a $+1$ to all the edges whose weight is either 3 or 4.\\

\begin{figure*}[t!]
\centering
\includegraphics[width=\textwidth]{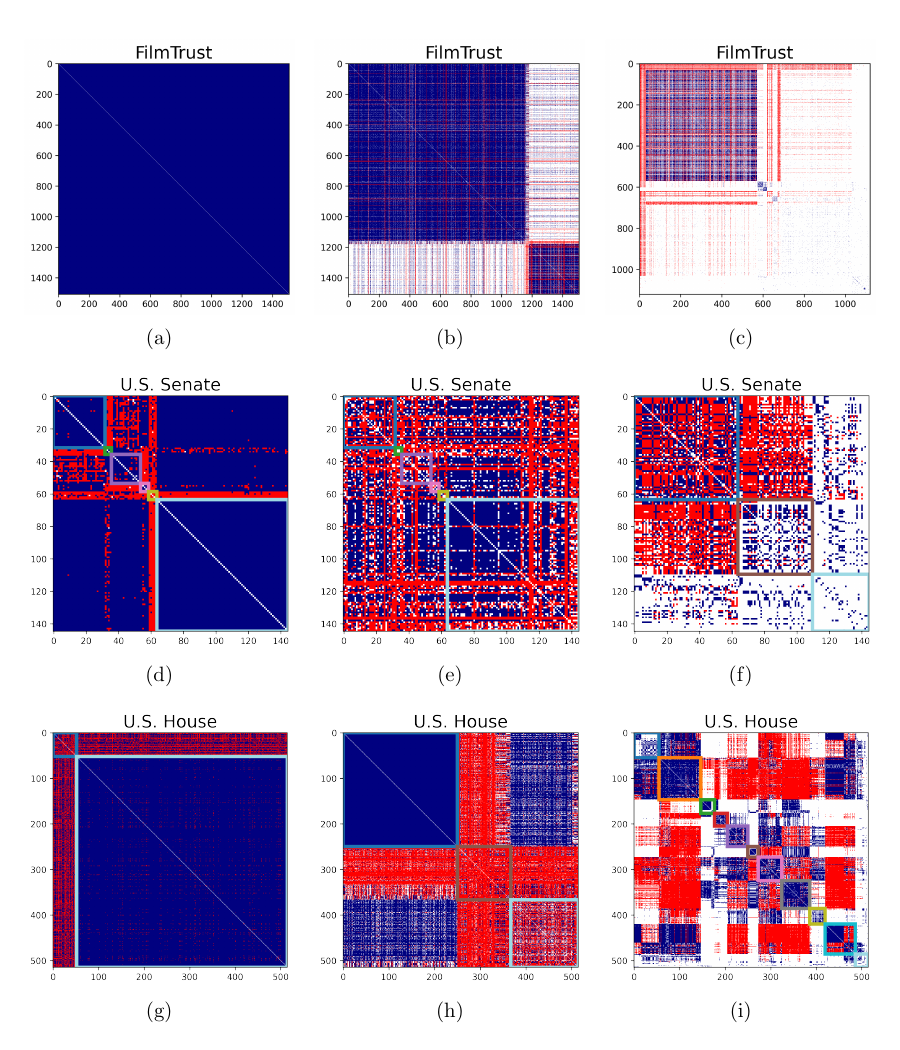}
\caption{Pictorial representation of the adjacency matrices of the projections of FilmTrust (panels (a)$-$(c)), U.S. Senate (panels (d)$-$(f)) and U.S. House of Representatives (panels (g)$-$(i)), obtained within the zero-inflated scheme, i.e. the na\"ive ones (panels (a), (d), (g)), the ones induced by the BiSRGM-FT (panels (b), (e), (h)) and the ones induced by the BiSCM-FT (panels (c), (f), (i)). Entries equal to $-1$ are coloured in red, entries equal to $0$ are coloured in white, entries equal to $+1$ are coloured in blue. The rows and columns of these adjacency matrices are re-ordered on the basis of the mesoscopic structures spotted by minimising BIC (see also Fig.~\ref{fig:7}).}
\label{fig:6}
\end{figure*}

From a purely empirical perspective, all the aforementioned datasets are characterised by a small link density $c=L/(N\cdot M)$: the connectance of \textit{U.S. Senate} and \textit{U.S. House of Representatives} amounts to $\simeq 0.1$ while the connectance of \textit{FilmTrust} amounts to $\simeq 0.01$. The percentages of positive and negative links are, instead, quite different: \textit{U.S. Senate} and \textit{U.S. House of Representatives} have $\simeq 55\%$ of positive links while \textit{FilmTrust} has $\simeq 80\%$ of positive links (see also Table~\ref{tab:iterative} in Appendix~\ref{AppD}).

\begin{figure*}
\centering
\includegraphics[width=\textwidth]{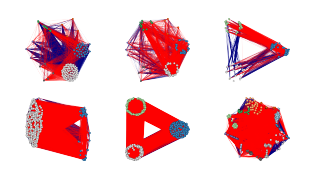}
\caption{Pictorial representation of the projections of U.S. Senate (top panels) and U.S. House of Representatives (bottom panels), obtained within the zero-inflated scheme, i.e. the na\"ive ones (left panels), the ones induced by the BiSRGM-FT (middle panels) and the ones induced by the BiSCM-FT (right panels). Entries equal to $-1$ are coloured in red, entries equal to $0$ are coloured in white, entries equal to $+1$ are coloured in blue. Minimising BIC lets the na\"ive projection of U.S. Senate (U.S. House of Representatives) to be partitioned into 6 (2) modules, the one induced by the BiSRGM-FT to be partitioned into 6 (3) modules and the one induced by the BiSCM-FT to be partitioned into 3 (11) modules.}
\label{fig:7}
\end{figure*}

\subsubsection{Na\"ive projection of real-world configurations}

The aforementioned configurations have been, first, na\"ively projected according to the recipes $a^\text{na\"ive}_{ij}=\text{sgn}[\underline{S_{ij}}]$, valid within the zero-deflated scheme, and $a^\text{na\"ive}_{ij}=\text{sgn}[S_{ij}]$, valid within the zero-inflated scheme.

When considering the unsigned case, the link density of the projection returned by the na\"ive approach is typically large~\cite{saracco2017inferring}. As shown in Fig.~\ref{fig:6}, this is especially true within the zero-inflated scheme - although the na\"ive projections of FilmTrust are both very dense. Let us also notice that the small link density of FilmTrust causes the na\"ive projection obtained within the zero-inflated scheme to be solely populated by positive links. For what concerns the na\"ive projection obtained within the zero-deflated scheme, instead, its large link density signals that the vast majority of pairs of nodes shares at least one V-motif (i.e. the absolute value of the difference between the number of `full' concordant and discordant motifs is at least 1). For what concerns U.S. Senate and U.S. House of Representatives, the na\"ive projections obtained within the zero-deflated scheme are already quite sparse, letting non-trivial mesoscopic patterns emerge - appreciable even by just looking at the adjacency matrices illustrated in the first column of Fig.~\ref{fig:5} (see also Table~\ref{table:1}).

\begin{figure*}
\centering
\includegraphics[width=\textwidth]{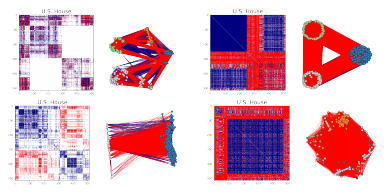}
\caption{Pictorial representation of the projections of U.S. House of Representatives, obtained within the zero-deflated scheme (first and second column) and the zero-inflated scheme (third and fourth column), induced by the BiSRGM-FT. Entries equal to $-1$ are coloured in red, entries equal to $0$ are coloured in white, entries equal to $+1$ are coloured in blue. The rows and columns of these adjacency matrices are re-ordered on the basis of the mesoscopic structures spotted by either minimising BIC (top panels) or the frustration $F$ (bottom panels). As the optimisation of $F$ is solely driven by signed membership, it returns a partition defined by \textit{i)} a smaller number of modules than those individuated by BIC when the majority of validated links is positive and \textit{ii)} a larger number of modules than those individuated by BIC when the majority of validated links is negative.}
\label{fig:8}
\end{figure*}

BIC minimisation confirms the presence of modules seemingly obeying the statistical variant of the relaxed balance theory (RBT)~\cite{gallo2024assessing,doreian2009partitioning}, as a non-negligible number of negative links is found not only \textit{between} clusters but \textit{within} clusters as well - the vast majority of links within the second block from the right-bottom angle of the (adjacency matrix of the) zero-deflated na\"ive projection of U.S. Senate is, in fact, negative - and a non-negligible number of positive links is found not only \textit{within} clusters but \textit{between} clusters as well - the vast majority of links between the first and the second block from the right-bottom angle of the (adjacency matrix of the) zero-deflated na\"ive projection of U.S. Senate is, in fact, positive.

As a last observation, let us stress that the large link density of the zero-inflated na\"ive projections of U.S. Senate and U.S. House of Representatives does not prevent BIC minimisation from detecting statistically significant mesoscopic structures: its sensitivity to the density of signed links~\cite{gallo2024assessing}, in fact, makes it capable of revealing the 6 different blocks constituting the zero-inflated na\"ive projection of U.S. Senate and the 2 different blocks constituting the zero-inflated na\"ive projection of U.S. House of Representatives, both depicted in Fig.~\ref{fig:7}.

\subsubsection{Validated projection of real-world configurations\\within the zero-deflated scheme}

Although summing motifs while retaining their own sign is, in a sense, enough to obtain sparse projections, the validation procedure proposed in this paper (also) aims at enhancing the identification of patterns encoding non-trivial information about the original structure.

For what concerns U.S. Senate, minimising BIC on the projection filtered via the global statistical benchmark named BiSRGM-FT refines the picture provided by minimising BIC on the na\"ive projection, confirming the presence of a larger number of modules - more precisely, 10 as shown in Fig.~\ref{fig:7}. Interestingly enough, all such modules are characterised by (a vast majority of) positive links, although positive links are found \textit{between} modules as well. Finally, minimising BIC on the projection filtered via the local statistical benchmark named BiSCM-FT returns a picture lying, somehow, halfway between the na\"ive one and the one filtered via the BiSRGM-FT, as 8 modules are, now, detected; contrarily to what has been revealed with the aid of the global filter, however, applying the local filter returns a picture where negative links are found \textit{within} modules as well.

The entire process is even more evident when considering U.S. House of Representatives, whose mesoscopic structure gets progressively resolved into an increasing number of increasingly negative modules (i.e. 2, 5, 6).

Overall, thus, our results confirm that the revealed modular structure seems to align better with the statistical variant of the RBT than with the statistical variant of its traditional counterpart (TBT), consistently across the different projections.

\subsubsection{Validated projection of real-world configurations\\within the zero-inflated scheme}

The considerations above are confirmed by the zero-inflated projections depicted in Fig.~\ref{fig:6}. For example, progressively filtering FilmTrust leads from a fully connected network to configurations whose connectance practically halves at each step. More in detail, applying the BiSRGM allows two different blocks to emerge - a result similar in spirit to the one that has been reported in~\cite{saracco2017inferring}, where the global statistical benchmark induces a rough partition of the system under consideration. The filtering induced by the BiSCM is, instead, more severe and qualitatively different: the projection obtained is, in fact, much sparser and populated by a comparable number of positive and negative links.

Interestingly, filtering U.S. Senate with the BiSRGM lets BIC identify the same number of modules characterising the na\"ive projection - they are 6 in both cases - but with a different arrangements of links; this is even more evident when considering the projection induced by the BiSCM, characterised by just three modules. On the contrary, filtering U.S. House of Representatives means rising (even substantially) the number of clusters while cutting half of the edges populating its na\"ive projection.

\section{Discussion}\label{sec:discussions}

A straightforward approach to project a bipartite network in the unsigned setting is that of comparing the number of neighbours shared by any two nodes $i$ and $j$, i.e.

\begin{equation}
V_{ij}=\sum_{\alpha=1}^Mb_{i\alpha}b_{j\alpha}=[\mathbf{B}\cdot\mathbf{B}^T]_{ij}=\mathbf{r}_i\cdot(\mathbf{r}_j)^T,
\end{equation}
with the one predicted under a chosen benchmark~\cite{saracco2017inferring}. In the signed setting, instead, two complementary approaches can be devised, treating the missing ties in a different way: while the zero-deflated projection scheme ignores them, the zero-inflated projection scheme accounts for them as well. Loosely speaking, the zero-deflated projection scheme returns configurations that are sparser than the configurations returned by the zero-inflated projection scheme although the latter are populated by a larger number of negative links; both schemes, however, enhance the detection of mesoscopic structures, the filtered projections being characterised by a larger number of modules than the na\"ive ones.

Additionally, we have compared the patterns revealed by minimising BIC with the ones revealed by minimising the frustration $F$, defined as

\begin{align}
F(\bm{\sigma})=L_\bullet^-+L_\circ^+
\end{align}
according to the traditional balance theory~\cite{gallo2024assessing}, i.e. counting the number of negative links within modules (indicated with a filled dot - see the first addendum) plus the number of positive links between modules (indicated with an empty dot - see the second addendum). The differences are highlighted in Fig.~\ref{fig:8}, showing how the optimisation of $F$ basically returns an oversimplified picture leading, for instance, to partition the projection of U.S. House of Representatives induced by the BiSRGM within the zero-inflated scheme into 32 modules. The explanation lies in the sensitivity of $F$ solely towards the \textit{signed membership}: as the \textit{signed density} is completely ignored, the groups of nodes dominated by negative links are split into singletons - and, by converse, the nodes connected by positive links are grouped together. Another example is provided by the projection of U.S. House of Representatives induced by the BiSRGM within the zero-deflated scheme: such a configuration is, now, partitioned into 2 modules, the signed membership leading the algorithm to disregard the (internal) hierarchical structure of these clusters. Similar results are found when considering the other kinds of projections.

A more quantitative comparison can be carried out upon calculating the \textit{Wallace}, \textit{Rand} and \textit{Jaccard Index}, that sum up the coefficients populating the confusion matrix to return three compact measures of similarity between partitions. As illustrated in Appendix~\ref{AppE}, the three aforementioned indices confirm that the mesoscale structures spotted by the BIC-based and the $F$-based recipes are more similar whenever a larger number of positive links populates a given projection. If, on the contrary, the latter is defined by a large number of negative links, the $F$-based recipe outputs many `false negatives', i.e. pairs of nodes that are separated \textit{solely} because found to connected by a negative link. Such a result sheds further light on the observation made in~\cite{gallo2024testing} and~\cite{gallo2024assessing} about the (potential) ambiguity concerning the variant of the balance theory best supported by the data: if the intuitive definition of modules as `densely connected groups of nodes' is extended to the signed case, then the recipe prescribing to minimise BIC should be preferred. If, on the other hand, one seeks the configuration best aligning with the TBT, then the recipe prescribing to minimise $F$ should be preferred - although not specifically designed to spot (statistically significant) mesoscale structures.\\

A very last observation concerns the `sign prediction' problem. In~\cite{singh2017measuring}, such an issue is addressed within the framework of the TBT, i.e. under the assumption that signed networks evolve towards balance: within such a context, sign prediction is carried out by combining the edge-based definition of Katz centrality with the minimisation of frustration. In~\cite{andres2023reconstructing}, instead, an approach based upon the ensemble of random graphs induced by the hypergeometric distribution is employed. Other methods are based upon machine learning techniques~\cite{dang2018link,derr2018signed,li2019integrating,huang2021signed,zhang2023contrastive,jiang2024sbgmn}. Our algorithm can, in a sense, be understood as implementing an unsupervised `white box' method for predicting the sign of a link: interestingly, the prescription based upon the signature of its endpoints - summarisable with the motto `a significantly large number of concordant motifs induces a $+1$ and a significantly large number of discordant motifs induces a $-1$' - formalises the tendency towards balance, advocated by other approaches, within the bipartite context.

\section{Methods}

\subsection{Formalism and basic quantities}\label{sec:formalism}

A binary undirected bipartite signed network is completely defined by its biadjacency matrix, i.e. a rectangular table $\mathbf{B}$ whose dimensions will be indicated with $M$ and $N$, $M$ being the number of nodes in the top layer (i.e. the number of columns of $\mathbf{B}$) and $N$ being the number of nodes in the bottom layer (i.e. the number of rows of $\mathbf{B}$).

Each edge can be \textit{positive}, \textit{negative} or \textit{missing}: since we will focus on binary networks, each edge will be `$+1$', `$-1$' or `$0$'. More formally, for any two nodes $i$ and $\alpha$, the corresponding entry of the biadjacency matrix will be assumed to read $b_{i\alpha}=-1,0,+1$.

To ease mathematical manipulations, let us define the three quantities reading

\begin{align}
b_{i\alpha}^-=[b_{i\alpha}=-1],\quad b_{i\alpha}^0=[b_{i\alpha}=0],\quad b_{i\alpha}^+=[b_{i\alpha}=+1],
\end{align}
where we have employed Iverson's brackets notation~\cite{gallo2024testing}. These new variables are mutually exclusive, satisfy the relationship $b_{i\alpha}^-+b_{i\alpha}^0+b_{i\alpha}^+=1$, $\forall\:i,\alpha$ and induce the three non-negative matrices $\mathbf{B}^+$, $\mathbf{B}^0$ and $\mathbf{B}^-$ obeying the relationships $\mathbf{B}=\mathbf{B}^+-\mathbf{B}^-$ and $|\mathbf{B}|=\mathbf{B}^++\mathbf{B}^-$.

In the following, it will turn out to be useful identifying a biadjacency matrix with the set of its rows, i.e. $\mathbf{B}\equiv\{\mathbf{r}_i\}_{i=1}^N$, and analogously for $\mathbf{B}^+$ and $\mathbf{B}^-$. Lastly, $\mathbf{B}^T$ indicates the transpose of the biadjacency matrix $\mathbf{B}$.

\subsection{Global connectivity and node degrees}

The number of positive and negative links, respectively, read

\begin{align}
L^+=\sum_{i=1}^N\sum_{\alpha=1}^Mb_{i\alpha}^+,\quad L^-=\sum_{i=1}^N\sum_{\alpha=1}^Mb_{i\alpha}^-;
\end{align}
analogously, the positive and negative degree of node $i$ are defined as

\begin{align}
k_i^+=\sum_{\alpha=1}^Mb_{i\alpha}^+,\quad k_i^-=\sum_{\alpha=1}^Mb_{i\alpha}^-
\end{align}
while the positive and negative degree of node $\alpha$ are defined as

\begin{align}
h_\alpha^+=\sum_{i=1}^Nb_{i\alpha}^+,\quad h_\alpha^-=\sum_{i=1}^Nb_{i\alpha}^-.
\end{align}

Naturally, $L^+=\sum_{i=1}^Nk_i^+=\sum_{\alpha=1}^Mh_\alpha^+$ and $L^-=\sum_{i=1}^Nk_i^-=\sum_{\alpha=1}^Mh_\alpha^-$. The advantage of adopting Iverson's brackets is that of ensuring that each quantity is, now, computed on a matrix with positive entries, i.e. is positive as well.

\subsection{The role of ambivalent patterns}

The approaches to obtain a projection of a bipartite, signed network considered so far rest upon the calculation of any two nodes similarity. Evaluating such a quantity leads to two related problems, i.e. how to treat ambivalent patterns and missing ties.\\

Ambivalence is introduced in~\cite{psycho1958robert} and defined as a `\textit{conjunction of positive and negative relations that are psychologically secondary or derived}'; in~\cite{cartwright1970ambivalence}, Cartwright and Harary suggest that `\textit{attitudes of ambivalence should be unstable, changing to positive or negative attitudes so as to satisfy the criteria of balance}'. In the bipartite setting we are considering here, ambivalence emerges in two, different cases: \textit{i)} whenever nodes $i$ and $j$ establish motifs whose signature is either $(+/-)$ or $(-/+)$; \textit{ii)} whenever nodes $i$ and $j$ establish the same number of $(+/+)$ and $(-/-)$ motifs. In both cases, devising a recipe to determine the sign of the link in the corresponding monopartite projection is not immediate.

In~\cite{schoch2021projecting}, two different recipes are proposed. The first one is based on matrix multiplication and prescribes to consider the projections induced by positive and negative links separately. The second one rely on a `vertex duplication' mechanism, according to which each node belonging to the layer of interest originates a positive copy, gathering the original positive links, and a negative copy, gathering the original negative links; these connections are treated as unsigned and the network is, then, projected. Finally, the sign of the edges populating the monopartite projection is restored according to a so-called `vertex contraction' rule.\\

In order to address the problem of the ambivalent patterns, let us define the `full' dyadic motifs reading

\begin{align}
V_{ij}^{++}&=\sum_{\alpha=1}^Mb_{i\alpha}^+b_{j\alpha}^+=[(\mathbf{B}^+)\cdot(\mathbf{B}^+)^T]_{ij}=\mathbf{r}^+_i\cdot(\mathbf{r}^+_j)^T,
\end{align}
that counts the number of nodes (belonging to the second layer) to which the nodes $i$ and $j$ (belonging to the first layer) are both connected via two positive links, and

\begin{align}
V_{ij}^{--}&=\sum_{\alpha=1}^Mb_{i\alpha}^-b_{j\alpha}^-=[(\mathbf{B}^-)\cdot(\mathbf{B}^-)^T]_{ij}=\mathbf{r}^-_i\cdot(\mathbf{r}^-_j)^T,
\end{align}
that counts the number of nodes (belonging to the second layer) to which the nodes $i$ and $j$ (belonging to the first layer) are both connected via two negative links. It is quite intuitive to ascribe them to the class of the so-called \textit{concordant motifs}, i.e. patterns capturing the `agreement' between any two users establishing them: our proposal is, thus, that of connecting any two users establishing a significantly large number of concordant motifs with a $+1$.

Analogously, it is quite intuitive to ascribe the expressions

\begin{align}
V_{ij}^{+-}&=\sum_{\alpha=1}^Mb_{i\alpha}^+b_{j\alpha}^-=[(\mathbf{B}^+)\cdot(\mathbf{B}^-)^T]_{ij}=\mathbf{r}^+_i\cdot(\mathbf{r}^-_j)^T
\end{align}
and

\begin{align}
V_{ij}^{-+}&=\sum_{\alpha=1}^Mb_{i\alpha}^-b_{j\alpha}^+=[(\mathbf{B}^-)\cdot(\mathbf{B}^+)^T]_{ij}=\mathbf{r}^-_i\cdot(\mathbf{r}^+_j)^T,
\end{align}
that count the number of nodes (belonging to the second layer) to which the nodes $i$ and $j$ (belonging to the first layer) are connected with a positive and a negative link, to the class of the so-called \textit{discordant motifs}, i.e. patterns capturing the `disagreement' between any two users establishing them: our proposal is, thus, that of connecting any two users establishing a significantly large number of discordant motifs with a $-1$. 

\subsection{The role of missing ties}

Let us, now, address the problem of the missing ties. To this aim, let us define the `partial' dyadic motifs reading

\begin{align}
V_{ij}^{00}&=\sum_{\alpha=1}^Mb_{i\alpha}^0b_{j\alpha}^0=[(\mathbf{B}^0)\cdot(\mathbf{B}^0)^T]_{ij}=\mathbf{r}^0_i\cdot(\mathbf{r}^0_j)^T,
\end{align}
that counts the number of nodes (belonging to the second layer) to which the nodes $i$ and $j$ (belonging to the first layer) are both not connected,

\begin{align}
V_{ij}^{0+}&=\sum_{\alpha=1}^Mb_{i\alpha}^0b_{j\alpha}^+=[(\mathbf{B}^0)\cdot(\mathbf{B}^+)^T]_{ij}=\mathbf{r}^0_i\cdot(\mathbf{r}^+_j)^T
\end{align}
and

\begin{align}
V_{ij}^{+0}&=\sum_{\alpha=1}^Mb_{i\alpha}^+b_{j\alpha}^0=[(\mathbf{B}^+)\cdot(\mathbf{B}^0)^T]_{ij}=\mathbf{r}^+_i\cdot(\mathbf{r}^0_j)^T,
\end{align}
that count the number of nodes (belonging to the second layer) to which node $i$ (node $j$) is not connected and node $j$ (node $i$) is connected with a positive link,

\begin{align}
V_{ij}^{0-}&=\sum_{\alpha=1}^Mb_{i\alpha}^0b_{j\alpha}^-=[(\mathbf{B}^0)\cdot(\mathbf{B}^-)^T]_{ij}=\mathbf{r}^0_i\cdot(\mathbf{r}^-_j)^T
\end{align}
and

\begin{align}
V_{ij}^{-0}&=\sum_{\alpha=1}^Mb_{i\alpha}^-b_{j\alpha}^0=[(\mathbf{B}^-)\cdot(\mathbf{B}^0)^T]_{ij}=\mathbf{r}^-_i\cdot(\mathbf{r}^0_j)^T,
\end{align}
that count the number of nodes (belonging to the second layer) to which node $i$ (node $j$) is not connected and node $j$ (node $i$) is connected with a negative link. While we are led to consider $V_{ij}^{00}$ as a signature of concordance (both nodes have no reason to connect to any of the nodes belonging to the opposite layer), we are also led to consider $V_{ij}^{0+}$, $V_{ij}^{+0}$, $V_{ij}^{0-}$, $V_{ij}^{-0}$ as a signature of discordance (one of the nodes has no reason to connect to any of the nodes belonging to the opposite layer while the other has\footnote{This is also logically coherent with the following situation: imagine $i$ likes item $\alpha$, $j$ has not a connection with it and $k$ dislikes it. Should $j$ be considered as agreeing with both $i$ and $k$, they also should for transitivity. But this is definitely not the case.}). The rightmost members indicate that the same numbers can be obtained by taking the scalar products of the rows, indexed by $i$ and $j$, of the biadjacency matrices $\mathbf{B}^+$, $\mathbf{B}^0$ and $\mathbf{B}^-$.\\

\begin{figure}[t!]
\centering
\includegraphics[width=0.45\textwidth]{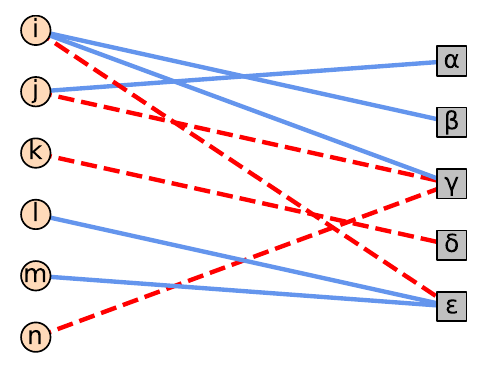}
\caption{Pictorial representation of a bipartite signed network with blue (positive) and red (negative) edges. The `full' dyadic motifs, forming the `bricks' of the zero-deflated projection scheme, are individuated by $ij$ and $in$ $(+/-)$, $il$ and $im$ $(-/+)$, $jn$ $(-/-)$ and $im$ $(+/+)$. Edges $i\beta$, $j\alpha$ and $k\delta$ do not contribute to any motif.}
\label{fig:1}
\end{figure}

In summary, while accounting for $V_{ij}^{++}$, $V_{ij}^{--}$, $V_{ij}^{+-}$ and $V_{ij}^{-+}$ will lead to a \textit{zero-deflated projection scheme}, where signs are solely determined by the `full' dyadic motifs, also considering $V_{ij}^{00}$, $V_{ij}^{0+}$, $V_{ij}^{+0}$, $V_{ij}^{0-}$ and $V_{ij}^{-0}$ will lead to a \textit{zero-inflated projection scheme}, where signs are determined by the `partial' dyadic motifs too. For a pictorial illustration of the aforementioned `full' and `partial' dyadic motifs, we refer the reader to Fig.~\ref{fig:1} and Fig.~\ref{fig:2}.

\subsection{A scheme for the statistical validation of bipartite signed networks}

Schematically, our validation algorithm works as follows:

\begin{itemize}
\item[$\mathbf{A.}$] focus on a specific pair of nodes belonging to the layer of interest, say $i$ and $j$, and measure their similarity (see Section~\ref{sec:step1});
\item[$\mathbf{B.}$] quantify the statistical significance of the measured similarity, with respect to a properly-defined benchmark, by computing the corresponding $p-$value, say $p_{ij}$ (see Section~\ref{sec:step2});
\item[$\mathbf{C.}$] repeat the step above for each pair of nodes;
\item[$\mathbf{D.}$] apply a multiple hypothesis testing procedure and connect the nodes $i$ and $j$ if and only if significantly similar (see Section~\ref{sec:step3}).
\end{itemize}

Let us stress that such a scheme is valid for both variants of our projection algorithm.

\subsection{Step \#1. Quantifying the similarity of any two nodes}\label{sec:step1}

Let us divide the next three sections in two subsections each: the first one will be devoted to devise a recipe for projecting a bipartite network that ignores the dyadic motifs constituted by, at least, one missing tie (the bipartite topology is considered `fixed' and only the signs of the `full' dyadic motifs are accounted for); the second one will be devoted to devise a recipe for projecting a bipartite network that accounts for the dyadic motifs constituted by, at least, one missing tie as well (the bipartite topology is considered `free' and the signs of both the `full' and the `partial' dyadic motifs are accounted for). To avoid confusion, the quantities defined within the first framework will be underlined.

\begin{figure}[t!]
\centering
\includegraphics[width=0.45\textwidth]{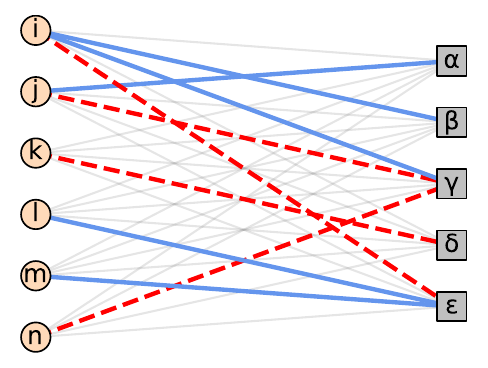}
\caption{Pictorial representation of a bipartite signed network with blue (positive), red (negative) and gray (missing) edges. The `full' dyadic motifs include $ij$ and $in$ $(+/-)$, $il$ and $im$ $(-/+)$, $jn$ $(-/-)$ and $im$ $(+/+)$. Additionally, node $i$ participates to `partial' motifs such as $(0/+)$, $(0/0)$, $(+/0)$ and $(0/-)$.}
\label{fig:2}
\end{figure}

\subsubsection{Zero-deflated projection scheme}\label{ssec:step1zdp_scheme}

The first step of our method prescribes to measure the degree of similarity of nodes $i$ and $j$. To this aim, let us consider the quantity named \textit{signature} and defined as:

\begin{align}
\underline{S_{ij}}&=\sum_{\underline{\alpha=1}}^{V_{ij}}b_{i\alpha}b_{j\alpha}\nonumber\\
&=\sum_{\underline{\alpha=1}}^{V_{ij}}[(b_{i\alpha}^+b_{j\alpha}^++b_{i\alpha}^-b_{j\alpha}^-)-(b_{i\alpha}^+b_{j\alpha}^-+b_{i\alpha}^-b_{j\alpha}^+)]\nonumber\\
&=\sum_{\underline{\alpha=1}}^{V_{ij}}(C_{ij\alpha}-D_{ij\alpha})\nonumber\\
&=\underline{C_{ij}}-\underline{D_{ij}}
\end{align}
where the sum runs over the `full' V-motifs\footnote{The symbol $\sum_{\underline{\alpha=1}}^{V_{ij}}(\dots)$ indicates that the sum runs over the connected pairs of nodes and is equivalent to $\sum_{\alpha=1}^M|b_{i\alpha}b_{j\alpha}|(\dots)$.}, i.e. $V_{ij}\equiv\sum_{\alpha=1}^M|b_{i\alpha}b_{j\alpha}|$. In words, the signature is the difference between two quantities, i.e. the \textit{concordance} of nodes $i$ and $j$, reading

\begin{align}
\underline{C_{ij}}&=\sum_{\underline{\alpha=1}}^{V_{ij}}C_{ij\alpha}=\sum_{\underline{\alpha=1}}^{V_{ij}}(b_{i\alpha}^+b_{j\alpha}^++b_{i\alpha}^-b_{j\alpha}^-)=\underline{V_{ij}}^{++}+\underline{V_{ij}}^{--}
\end{align}
and counting the number of `full' concordant motifs, and the \textit{discordance} of nodes $i$ and $j$, reading

\begin{align}
\underline{D_{ij}}&=\sum_{\underline{\alpha=1}}^{V_{ij}}D_{ij\alpha}=\sum_{\underline{\alpha=1}}^{V_{ij}}(b_{i\alpha}^+b_{j\alpha}^-+b_{i\alpha}^-b_{j\alpha}^+)=\underline{V_{ij}}^{+-}+\underline{V_{ij}}^{-+}
\end{align}
and counting the number of `full' discordant motifs. As Fig.~\ref{fig:1} shows, the pairs of nodes establishing motifs that are accounted for in the zero-deflated scheme are $ij$ and $in$, establishing a $(+/-)$ motif; $il$ and $im$, establishing a $(-/+)$ motif; $jn$, establishing a $(-/-)$ motif; $lm$, establishing a $(+/+)$ motif.

A na\"ive way of projecting a bipartite signed network would prescribe to stop here and apply the sign function to the signature, hence connecting nodes $i$ and $j$ with a positive link if $\underline{S_{ij}}>0$, i.e. $\underline{C_{ij}}>\underline{D_{ij}}$, and with a negative link if $\underline{S_{ij}}<0$, i.e. $\underline{C_{ij}}<\underline{D_{ij}}$. More compactly,

\begin{equation}
a^\text{na\"ive}_{ij}=\text{sgn}[\underline{S_{ij}}].
\end{equation}

\subsubsection{Zero-inflated projection scheme}

Let us, now, measure the degree of similarity of nodes $i$ and $j$ within the zero-inflated projection scheme. To this aim, let us consider the novel definition of signature reading

\begin{align}
S_{ij}=\sum_{\alpha=1}^M(C_{ij\alpha}-D_{ij\alpha})=C_{ij}-D_{ij}
\end{align}
where the concordance between nodes $i$ and $j$, now, reads

\begin{align}
C_{ij}=\sum_{\alpha=1}^M C_{ij\alpha}=V_{ij}^{++}+V_{ij}^{--}+V_{ij}^{00}
\end{align}
and the discordance between nodes $i$ and $j$, now, reads

\begin{align}
D_{ij}&=\sum_{\alpha=1}^M D_{ij\alpha}\nonumber\\
&=V_{ij}^{+-}+V_{ij}^{-+}+V_{ij}^{0+}+V_{ij}^{+0}+V_{ij}^{0-}+V_{ij}^{-0}.
\end{align}

As Fig.~\ref{fig:2} shows, the pairs of nodes establishing `full' motifs are $ij$ and $in$, establishing a $(+/-)$ motif; $il$ and $im$, establishing a $(-/+)$ motif; $jn$, establishing a $(-/-)$ motif; $lm$, establishing a $(+/+)$ motif. Moreover, node $i$ establishes a $(0/+)$ motif with node $j$ via node $\alpha$, a $(0/0)$ motif with nodes $k$, $l$, $m$ and $n$ via node $\alpha$, a $(+/0)$ motif with any other node via node $\beta$, a $(0/-)$ motif with node $k$ via node $\delta$, etc.

A na\"ive way of projecting a bipartite signed network would prescribe to stop here and apply the sign function to the signature, connecting nodes $i$ and $j$ with a positive link if $S_{ij}>0$, i.e. $C_{ij}>D_{ij}$, and with a negative link if $S_{ij}<0$, i.e. $C_{ij}<D_{ij}$. More compactly,

\begin{align}
a^\text{na\"ive}_{ij}=\text{sgn}[S_{ij}].
\end{align}
Notice that such a na\"ive projection would be denser than the na\"ive one induced by the zero-deflated definition of signature, the only possibility of observing $a^\text{na\"ive}_{ij}=0$ being that of having $C_{ij}=D_{ij}$.

\subsection{Step \#2. Quantifying the statistical significance of similarity}\label{sec:step2}

\subsubsection{Zero-deflated projection scheme}

The second step of our method prescribes to evaluate the statistical significance of our nodes similarity. To this aim, let us find the probability distribution obeyed by $\underline{S_{ij}}$, after noticing that

\begin{equation}
-V_{ij}\leq\underline{S_{ij}}\leq V_{ij}
\end{equation}
where $\underline{S_{ij}}=-V_{ij}$ if $\underline{C_{ij}}=0$ (i.e. if each `full' V-motif is composed by a $-1$ and a $+1$) and $\underline{S_{ij}}=V_{ij}$ if $\underline{D_{ij}}=0$ (i.e. if each `full' V-motif is composed by either two $-1$s or two $+1$s).\\

Let us, now, keep the formalism as general as possible and treat links as independent non-identically distributed (i.n.i.d.) random variables. This amounts at considering the finite scheme\footnote{Such a notation, introduced by Khintchine in \emph{Mathematical Foundations of Information Theory}~\cite{khintchine1957mathematical}, compactly represents a discrete probability distribution, by listing its support on the first row and the related probability coefficients on the second row.}

\begin{equation}\label{eq:nmheterofixed}
b_{i\alpha}\sim
\begin{pmatrix}
-1  & +1\\
1-p_{i\alpha}^+ & p_{i\alpha}^+
\end{pmatrix}
\end{equation}
$\forall\:i,\alpha$ such that $|b_{i\alpha}|=1$, further inducing

\begin{align}\label{eq:nmheterofixed2}
b_{i\alpha}b_{j\alpha}&\sim
\begin{pmatrix}
-1 & +1\\
p_{i\alpha}^++p_{j\alpha}^+-2p_{i\alpha}^+p_{j\alpha}^+ & 1-p_{i\alpha}^+-p_{j\alpha}^++2p_{i\alpha}^+p_{j\alpha}^+
\end{pmatrix}\nonumber\\
&=\begin{pmatrix}
-1 & +1\\
1-q_{ij\alpha}^+ & q_{ij\alpha}^+
\end{pmatrix}
\end{align}
$\forall\:i,j,\alpha$ such that $|b_{i\alpha}b_{j\alpha}|=1$.

Since $\underline{S_{ij}}$ is the sum of i.n.i.d. Bernoulli random variables, the probability distribution obeyed by it is the Poisson-binomial

\begin{align}
P(\underline{S_{ij}}=s)&=\sum_{C_k\in\mathcal{C}_k}\left[\prod_{\nu\in C_k}q_{ij\nu}^+\prod_{\tau\notin C_k}(1-q_{ij\tau}^+)\right]
\end{align}
where $\mathcal{C}_k$ is the set of all possible $k$-tuples of which $\nu$ and $\tau$ are instances. As we will see in Section~\ref{sec:choosing}, this is precisely the case of a benchmark enforcing linear \emph{local} constraints.

The formula above simplifies in the case of a benchmark enforcing linear \emph{global} constraints, the links becoming independent identically distributed (i.i.d.) random variables and the Poisson-binomial above reducing to a binomial.

For more details about the benchmarks, see Sections~\ref{homofixed}, ~\ref{heterofixed} and Appendix~\ref{AppA}.

\begin{figure*}[t!]
\centering
\includegraphics[width=\textwidth]{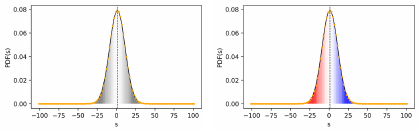}
\caption{Probability distribution of the signature (\textcolor{forest}{$\bullet$}) and its Gaussian approximation (\textcolor{orange}{$\bullet$}). The left panel provides a graphical answer to the question \textit{is the empirical value of the signature significantly different from the one expected under the chosen benchmark?} while the right panel provides a graphical answer to the question \textit{is the deviation negative (hence, the signature is significantly smaller than expected) or positive (hence, the signature is significantly larger than expected)?}, the red (blue) area corresponding to the region of validation of the negative (positive) links.}
\label{fig:3}
\end{figure*}

\subsubsection{Zero-inflated projection scheme}\label{ssec:step2zdp_scheme}

Let us, now, find the probability distribution obeyed by $S_{ij}$, after noticing that

\begin{align}
-M\leq S_{ij}\leq M
\end{align}
where $S_{ij}=-M$ if $C_{ij}=0$ (i.e. if each V-motif is composed by a $-1$ and a $+1$) and $S_{ij}=M$ if $D_{ij}=0$ (i.e. if each V-motif is composed by either two $-1$s or two $+1$s).\\

As in the previous subsection, let us treat links link as i.n.i.d. random variables. Within the zero-inflated projection scheme, this amounts at replacing Eq.~\ref{eq:nmheterofixed} with

\begin{equation}\label{eq:nmheterofree}
b_{i\alpha}\sim
\begin{pmatrix}
-1 & 0 & +1\\
p_{i\alpha}^- & p_{i\alpha}^0 & p_{i\alpha}^+
\end{pmatrix}
\end{equation}
$\forall\:i,\alpha$, further inducing

\begin{align}
C_{ij\alpha}-D_{ij\alpha}\sim
\begin{pmatrix}
-1 & +1\\
1-q_{ij\alpha}^+ & q_{ij\alpha}^+
\end{pmatrix}
\end{align}
with

\begin{align}
q_{ij\alpha}^+&=q_{ij\alpha}^{++}+q_{ij\alpha}^{--}+q_{ij\alpha}^{00},\\
1-q_{ij\alpha}^+&=q_{ij\alpha}^{+-}+q_{ij\alpha}^{-+}+q_{ij\alpha}^{0+}+q_{ij\alpha}^{+0}+q_{ij\alpha}^{0-}+q_{ij\alpha}^{-0}
\end{align}
where, for example, the coefficient $q_{ij\alpha}^{++}$ induces the finite scheme

\begin{align}
b_{i\alpha}^+b_{j\alpha}^+&\sim
\begin{pmatrix}
0 & +1\\
1-p_{i\alpha}^+p_{j\alpha}^+ & p_{i\alpha}^+p_{j\alpha}^+
\end{pmatrix}=\begin{pmatrix}
0 & +1\\
1-q_{ij\alpha}^{++} & q_{ij\alpha}^{++}
\end{pmatrix}
\end{align}
and analogously for the others.

As a consequence, $S_{ij}$ becomes a sum of i.n.i.d. Bernoulli random variables obeying the Poisson-binomial reading

\begin{align}
P(S_{ij}=s)&=\sum_{C_k\in\mathcal{C}_k}\left[\prod_{\nu\in C_k}q_{ij\nu}^+\prod_{\tau\notin C_k}(1-q_{ij\tau}^+)\right]
\end{align}
where $\mathcal{C}_k$ is the set of all possible $k$-tuples of which $\nu$ and $\tau$ are instances. As in the previous subsection, if the benchmark induced by linear local constraints is replaced by the benchmark induced by linear global constraints, the links become i.i.d. random variables and the Poisson-binomial reduces to a binomial.

For more details about the benchmarks, see Sections~\ref{homofree}, ~\ref{heterofree} and Appendix~\ref{AppB}.

\subsubsection{Quantifying the statistical significance of similarity}

Once we have calculated the distribution for each pair of nodes belonging to the layer of interest, we have to calculate the statistical significance of the empirical signature: as we have a signed quantity, we need both tails of such a distribution to carry out what is known as \textit{two-sided test of hypothesis}. In what follows we will focus on $\underline{S_{ij}}$ but the same considerations hold true for $S_{ij}$.

Basically, we need to answer the two related questions \textit{i) is the empirical value of the signature significantly different from the one expected under the chosen benchmark?} and \textit{ii) if so, is the deviation negative (hence, the signature is significantly smaller than expected) or positive (hence, the signature is significantly larger than expected)?}\\

The first question can be answered upon calculating the two-sided $p-$value reading

\begin{equation}\label{p2}
p_{ij}=2\cdot\min\left\{F(\underline{S_{ij}^*}), 1-F(\underline{S_{ij}^*})\right\}
\end{equation}
with $S_{ij}^*$ being the empirical value of the signature and $F$ being the cumulative distribution function, defined as $F(\underline{S_{ij}^*})=\sum_{x\leq\underline{S_{ij}^*}}P(\underline{S_{ij}}=x)$: such a number evaluates the probability of observing a deviation from the expected value in either directions.

The second question can be answered upon calculating the sign of such a deviation, by determining if either

\begin{equation}
F(\underline{S_{ij}^*})<1-F(\underline{S_{ij}^*})
\end{equation}
or

\begin{equation}
F(\underline{S_{ij}^*})>1-F(\underline{S_{ij}^*})
\end{equation}
holds true. In the first case, $F(\underline{S_{ij}^*})<1/2$, the empirical value of the signature is smaller than the median of the distribution and the deviation is negative; in the second case, $F(\underline{S_{ij}^*})>1/2$, the empirical value of the signature is larger than the median of the distribution and the deviation is positive. Upon indicating the threshold individuated by the multiple hypothesis testing procedure with $p_{th}$\footnote{It represents the largest $p-$value that satisfies the False Discovery Rate rejection criterion and is defined as $p_{th}=\hat{i}t/|H|$, with $\hat{i}$ being the largest integer satisfying the condition $\text{$p-$value}_{\hat{i}}\le p_{th}$, $t$ is the single-test significance level and $|H|$ is the total number of tested hypotheses (see also Appendix~\ref{App:FDR}).}, the following cases can be met:

\begin{itemize}
\item[$\bullet$] the two conditions $F(\underline{S_{ij}^*})<1/2$ and $p_{ij}\leq p_{th}$ indicate that nodes $i$ and $j$ have established a number of `full' discordant motifs that is so large to induce a significantly negative signature - and, potentially, a negative link in the projection ($a_{ij}=-1$);
\item[$\bullet$] the two conditions $F(\underline{S_{ij}^*})>1/2$ and $p_{ij}\leq p_{th}$ indicate that nodes $i$ and $j$ have established a number of `full' concordant motifs that is so large to induce a significantly positive signature - and, potentially, a positive link in the projection ($a_{ij}=+1$);
\item[$\bullet$] the condition $p_{ij}>p_{th}$ individuates a value of the signature (i.e. of `full' concordant/discordant motifs) that is \textit{compatible} with the one predicted by the chosen benchmark: stated otherwise, the empirical number of motifs could have been observed in configurations generated by the benchmark itself - hence, inducing a null link in the projection ($a_{ij}=0$).
\end{itemize}

For a graphical representation of our validation procedure, see Fig.~\ref{fig:3}.

\subsection{Step \#3. Validating the projection}\label{sec:step3}

The second step of our method returns a symmetric matrix of $p-$values. Individuating the ones associated with the hypotheses to be actually rejected requires a procedure to deal with the comparison of multiple hypotheses at the same time. In very general terms, a threshold has to be set: if the specific $p-$value is smaller than such a threshold, the associated event is interpreted as statistically significant and the corresponding nodes are connected in the projection.

In the present paper, we apply the so-called False Discovery Rate (FDR) procedure~\cite{benjamini1995controlling}, allowing one to control for the expected number of `false discoveries' (i.e. incorrectly rejected null hypotheses or incorrectly validated links), irrespectively of the independence of the hypotheses tested. 

For more details about the FDR procedure, see Appendix~\ref{App:FDR}.

\subsection{Maximum entropy benchmarks for bipartite signed networks}\label{sec:choosing}

The second step of our method prescribes to quantify the statistical significance of the similarity of any two nodes $i$ and $j$. To this aim, a statistical benchmark is needed. A natural choice leads to the adoption of models belonging to the class of the so-called Exponential Random Graphs, induced by the constrained maximisation of the Shannon entropy~\cite{jaynes1957information,park2004statistical,Squartini2011a,Squartinia,saracco2015randomizing,Cimini2018a}

\begin{equation}
S=-\sum_{\mathbf{B}\in\mathbb{B}}P(\mathbf{B})\ln P(\mathbf{B}),
\end{equation}
the sum running over a properly defined ensemble of configurations. Within such a framework, the generic bipartite network $\mathbf{B}$ is assigned the probability

\begin{align}
P(\mathbf{B})=\frac{e^{-H(\bm{\theta},\bm{C}(\mathbf{B}))}}{Z(\bm{\theta})}
\end{align}
whose value is determined by the vector $\bm{C}(\mathbf{B})=\{C_i(\mathbf{B})\}$ of topological constraints via the Hamiltonian reading $H(\bm{\theta},\bm{C}(\mathbf{B}))=\sum_i\theta_iC_i(\mathbf{B})$. In what follows, we will employ linear constraints: beside allowing our optimisation problem to be analytically solved in these cases - $P(\mathbf{B})$ can be written in a factorised form, i.e. as a product of pair-specific probability coefficients - they individuate the `right' amount of information to be accounted for to obtain a projection: since we are interested in evaluating the statistical significance of co-occurrences, one must discount the tendency of nodes to establish connections with many/few neighbours.

In order to determine the unknown parameters $\bm{\theta}$, the likelihood-maximisation recipe can be adopted: given an observed biadjacency matrix $\mathbf{B}^*$, it translates into solving the system of equations

\begin{equation}
\langle C_i\rangle(\bm{\theta})=\sum_{\mathbf{B}\in\mathbb{B}}P(\mathbf{B})C_i(\mathbf{B})=C_i(\mathbf{B}^*),\quad\forall\:i
\end{equation}
which prescribes to equate each ensemble average, e.g. $\langle C_i\rangle(\bm{\theta})$, to its observed counterpart, i.e. $C_i(\mathbf{B}^*)$~\cite{garlaschelli2008,Squartini2011a,Squartinia,saracco2015randomizing,Cimini2018a}.

\subsubsection{Zero-deflated projection scheme}

The first two benchmarks determine the coefficients of the two variants of the finite scheme defined in Eq.~\ref{eq:nmheterofixed}. As already stressed, dyadic motifs constituted by, at least, one missing tie are ignored, here: the bipartite topology is, thus, considered `fixed'. Upon indicating the total number of connected pairs of nodes with $L=\sum_{i=1}^N\sum_{\alpha=1}^M|b_{i\alpha}|=L^++L^-$ and considering that any node pair can be either positively or negatively connected, the ensemble is constituted by $|\mathbb{B}|=2^L$ possible configurations in both cases.\\

\paragraph*{Fixed-topology Bipartite Signed Random Graph Model.}\label{homofixed} The Fixed-topology Bipartite Signed Random Graph Model (BiSRGM-FT) is defined by two global constraints, i.e. the total number of positive and negative links: the corresponding Hamiltonian, thus, reads

\begin{equation}\label{eq:H_BiSRM_FT}
H(\bm{\theta},\mathbf{B})=\beta' L^+(\mathbf{B})+\gamma' L^-(\mathbf{B}).
\end{equation}

Keeping the topology of the network under analysis fixed while (solely) randomising the edge signs implies that the role of random variables is played by the entries of the biadjacency matrix corresponding to the connected pairs of nodes, i.e. the ones for which $|b_{i\alpha}|=1$. Let us, however, notice that the Hamiltonian in Eq.~\ref{eq:H_BiSRM_FT} can be re-written as

\begin{align}
H(\bm{\theta},\mathbf{B})&=\beta' L^+(\mathbf{B})+\gamma' (L-L^+)(\mathbf{B})\\
&=(\beta'-\gamma')L^+(\mathbf{B})+\gamma' L(\mathbf{B})
\end{align}
and further simplified into

\begin{align}
H(\bm{\theta},\mathbf{B})=\beta L^+(\mathbf{B})
\end{align}
where the constant term has been dropped, as it does not affect the computation of the probability coefficients, and the only relevant Lagrange multiplier has been re-named~\cite{hao2024proper}. The generic entry, thus, obeys

\begin{align*}
P(b_{i\alpha}=+1)&=\frac{e^{-\beta}}{1+e^{-\beta}}=p^+,\\
P(b_{i\alpha}=-1)&=\frac{1}{1+e^{-\beta}}=1-p^+;
\end{align*}
in words, each entry satisfying $|b_{i\alpha}|=1$ obeys a Bernoulli distribution whose probability coefficients are determined by the imposed constraints: each existing link is assigned a `plus one' with probability $p^+$ and a `minus one' with probability $p^-$.

To employ the BiSRGM-FT for studying real-world networks, the parameters that define it need to be properly tuned. More specifically, one needs to ensure

\begin{align}
\langle L^+\rangle_\text{BiSRGM-FT}&=L^+(\mathbf{B}^*)
\end{align}
with the symbol $\mathbf{B}^*$ indicating the empirical network under analysis. The maximisation of the likelihood function $\mathcal{L}_\text{BiSRGM-FT}\equiv\ln P_\text{BiSRGM-FT}(\mathbf{B}^*)$ with respect to the unknown parameters that define it leads us to find

\begin{align}
p^+&=L^+(\mathbf{B}^*)/L(\mathbf{B}^*).
\end{align}

For an alternative, yet equivalent, resolution of the problem, see Appendix~\ref{AppC}.\\

\paragraph*{Fixed-topology Bipartite Signed Configuration Model.}\label{heterofixed} The Fixed-topology Bipartite Signed Configuration Model (BiSCM-FT), instead, is defined by local constraints, i.e. the positive and negative degree sequences: the corresponding Hamiltonian, thus, reads

\begin{align}\label{eq:H_biscm_ft}
H(\bm{\theta},\mathbf{B})=&\sum_{i=1}^N[\beta'_ik_i^+(\mathbf B)+\gamma'_ik_i^-(\mathbf B)]\nonumber\\
&+\sum_{\alpha=1}^M[\delta'_\alpha h_\alpha^+(\mathbf B)+\eta'_\alpha h_\alpha^-(\mathbf B)];
\end{align}
again, the role of random variables is played by the entries of the biadjacency matrix corresponding to the connected pairs of nodes and the Hamiltonian in Eq.~\ref{eq:H_biscm_ft} can be further simplified into

\begin{align}
H(\bm{\theta},\mathbf{B})=&\sum_{i=1}^N[\beta'_ik_i^+(\mathbf B)+\gamma'_i(k_i-k_i^+(\mathbf B))]\\
&+\sum_{\alpha=1}^M[\delta'_\alpha h_\alpha^+(\mathbf B)+\eta'_\alpha (h_\alpha-h_\alpha^+(\mathbf B))]\\
=&\sum_{i=1}^N\beta_ik_i^+(\mathbf B)+\sum_{\alpha=1}^M\delta_\alpha h_\alpha^+(\mathbf B)
\end{align}
with obvious meaning of the symbols~\cite{hao2024proper}. The generic entry, now, obeys

\begin{align}
P(b_{i\alpha}=+1)&=\frac{e^{-(\beta_i+\delta_\alpha)}}{1+e^{-(\beta_i+\delta_\alpha)}}=p_{i\alpha}^+,\\
P(b_{i\alpha}=-1)&=\frac{1}{1+e^{-(\beta_i+\delta_\alpha)}}=1-p_{i\alpha}^+;
\end{align}
in words, given any two connected nodes $i$ and $\alpha$, their link is assigned a $+1$ with probability $p_{i\alpha}^+$ and a $-1$ with probability $p_{i\alpha}^-$.

To tune the parameters defining the BiSCM-FT, we can maximise the likelihood function $\mathcal{L}_\text{BiSCM-FT}\equiv\ln P_\text{BiSCM-FT}(\mathbf{B}^*)$ with respect to the unknown parameters that define it: such a recipe leads us to find

\begin{align}\label{eq:maxlik_biscm_ft}
\langle k_i^+\rangle_\text{BiSCM-FT}&=k_i^+(\mathbf{B}^*),\quad\forall\:i,\\
\langle h_\alpha^+\rangle_\text{BiSCM-FT}&=h_\alpha^+(\mathbf{B}^*),\quad\forall\:\alpha.
\end{align}

The system above can be solved only numerically, along the guidelines provided in~\cite{vallarano2021}.

For an alternative, yet equivalent, resolution of the problem, see Appendices D and E. Here, we have solved it by employing the SIMONA Matlab-coded package, available at this \href{https://it.mathworks.com/matlabcentral/fileexchange/167426-signed-models-for-network-analysis}{URL}.

\subsubsection{Zero-inflated projection scheme}

The other two benchmarks determine the coefficients of the two variants of the finite scheme defined in Eq.~\ref{eq:nmheterofree}. As already stressed, dyadic motifs constituted by, at least, one missing tie are accounted for, here: the bipartite topology is, thus, considered `free'. Since the total number of node pairs is $N\cdot M$ and any node pair can be positively connected, negatively connected or disconnected, the ensemble is constituted by $|\mathbb{B}|=3^{N\cdot M}$ possible configurations in both cases.\\

\paragraph*{Free-topology Bipartite Signed Random Graph Model.}\label{homofree} The Free-topology Bipartite Signed Random Graph Model (BiSRGM)~\cite{gallo2024testing} is induced by the same Hamiltonian inducing its fixed-topology counterpart, i.e. $H(\bm{\theta},\mathbf{B})=\beta L^+(\mathbf{B})+\gamma L^-(\mathbf{B})$ but treating the topology as `free' does not allow for any simplification. The generic entry obeys

\begin{align}
P(b_{i\alpha}=+1)&=\frac{e^{-\beta}}{1+e^{-\beta}+e^{-\gamma}}=p^+,\\
P(b_{i\alpha}=-1)&=\frac{e^{-\gamma}}{1+e^{-\beta}+e^{-\gamma}}=p^-
\end{align}
and $p^0\equiv1-p^--p^+$. In words, $b_{i\alpha}$ obeys a generalised Bernoulli distribution whose probability coefficients are determined by the imposed constraints. Each positive link appears with probability $p^+$, each negative link appears with probability $p^-$ and each missing link has a probability $p^0$.

The maximisation of the likelihood function $\mathcal{L}_\text{BiSRGM}=\ln P_\text{BiSRGM}(\mathbf{B}^*)$ with respect to the unknown parameters that define it leads us to find

\begin{align}
p^+&=L^+(\mathbf{B}^*)/(N\cdot M),\\
p^-&=L^-(\mathbf{B}^*)/(N\cdot M)
\end{align}
and $p^0\equiv 1-p^--p^+$.

For more details, see Appendix~\ref{AppC}.\\

\paragraph*{Free-topology Bipartite Signed Configuration Model.}\label{heterofree} The Free-topology Bipartite Signed Configuration Model (BiSCM)~\cite{gallo2024testing} is induced by the same Hamiltonian inducing its fixed-topology counterpart, i.e. $H(\bm{\theta},\mathbf{B})=\sum_{i=1}^N[\beta_ik_i^+(\mathbf B)+\gamma_ik_i^-(\mathbf B)]+\sum_{\alpha=1}^M[\delta_\alpha h_\alpha^+(\mathbf B)+\eta_\alpha h_\alpha^-(\mathbf B)]$ but, as for the BiSRGM, it cannot be further simplified. The generic entry, thus, reads

\begin{align}
P(b_{i\alpha}=+1)&=\frac{e^{-(\beta_i+\delta_\alpha)}}{1+e^{-(\beta_i+\delta_\alpha)}+e^{-(\gamma_i+\eta_\alpha)}}=p_{i\alpha}^+,\\
P(b_{i\alpha}=-1)&=\frac{e^{-(\gamma_i+\eta_\alpha)}}{1+e^{-(\beta_i+\delta_\alpha)}+e^{-(\gamma_i+\eta_\alpha)}}=p_{i\alpha}^-
\end{align}
and $p_{i\alpha}^0\equiv1-p_{i\alpha}^--p_{i\alpha}^+$. In words, $b_{i\alpha}$ obeys a generalised Bernoulli distribution whose probability coefficients are determined by the imposed constraints. Given any two nodes $i$ and $\alpha$, they are connected by a positive link with probability $p_{i\alpha}^+$, by a negative link with probability $p_{i\alpha}^-$ and are disconnected with probability $p_{i\alpha}^0$.

The maximisation of the likelihood function $\mathcal{L}_\text{BiSCM}=\ln P_\text{BiSCM}(\mathbf{B}^*)$ with respect to the unknown parameters that define it leads us to find

\begin{align}
k_i^+(\mathbf{B}^*)&=\langle k_i^+\rangle,\quad\forall\:i,\\
k_i^-(\mathbf{B}^*)&=\langle k_i^-\rangle,\quad\forall\:i,\\
h_\alpha^+(\mathbf{B}^*)&=\langle h_\alpha^+\rangle,\quad\forall\:\alpha,\\
h_\alpha^-(\mathbf{B}^*)&=\langle h_\alpha^-\rangle,\quad\forall\:\alpha.
\end{align}

The system above can be solved only numerically, along the guidelines provided in~\cite{vallarano2021}.

For more details, see Appendices D and E. Here, we have solved it by employing the SIMONA Matlab-coded package, available at this \href{https://it.mathworks.com/matlabcentral/fileexchange/167426-signed-models-for-network-analysis}{URL}.

Let us conclude this section by mentioning that BiSCM can be obtained as a special case of the Bipartite Score Configuration Model presented in~\cite{becatti2019}.

\section{Data availability}

Data concerning \textit{U.S. Senate} and \textit{U.S. House of Representatives} are described in~\cite{derr2019balance} and can be found at the address \url{https://www.govtrack.us/}. Data concerning \textit{FilmTrust} is described in~\cite{filmtrust} and can be found at the address \url{http://konect.cc/networks/librec-filmtrust-ratings/}.

\section{Code availability}

We released a Matlab-coded package that implements all the probabilistic models for binary undirected bipartite signed network: its name is SIMONA, an acronym standing for `Signed Models for Network Analysis', and is freely downloadable at this \href{https://it.mathworks.com/matlabcentral/fileexchange/167426-signed-models-for-network-analysis}{URL}.

\section{Acknowledgments}

AG and TS acknowledge support from the projects `SoBigData.it - Strengthening the Italian RI for Social Mining and Big Data Analytics' - IR0000013 - CUP B53C22001760006, financed by European Union - Next Generation EU - National Recovery and Resilience Plan (Piano Nazionale di Ripresa e Resilienza, PNRR) - M4C2 I.3.1; `Reconstruction, Resilience and Recovery of Socio-Economic Networks' RECON-NET EP\_FAIR\_005 - PE0000013 `FAIR' - PNRR M4C2 Investment 1.3, financed by European Union - Next Generation EU; PRIN 2022 2022MTBB22 `RE-Net: Reconstructing economic networks: from physics to machine learning and back', financed by the European Union - Next Generation EU, M4C2 Inv. 1.1 CUP D53D23002330006; PRIN 2022 PNRR P2022E93B8 `C2T - From Crises to Theory: towards a science of resilience and recovery for economic and financial systems', financed by the European Union - Next Generation EU, M4C2 Inv. 1.1, CUP: D53D23019330001. FS was partially supported by the project `CODE – Coupling Opinion Dynamics with Epidemics', financed under PNRR Mission 4 `Education and Research' - Component C2 - Investment 1.1 - Next Generation EU `Fund for National Research Program and Projects of Significant National Interest' PRIN 2022 PNRR, grant code P2022AKRZ9, CUP B53D23026080001.

\section{Author contributions}

Study conception and design: AG, FS, TS. Data collection: AG. Analysis and interpretation of results: AG, FS, TS. Draft manuscript preparation: AG, FS, TS.

\section{Competing Interests}

The authors declare no competing interests.

\bibliography{bibmain.bib}

\clearpage

\onecolumngrid

\appendix

\section{The distribution of the signature in the zero-deflated scheme}\label{AppA}

Let us, first, remind that such a scheme devises a recipe for projecting a bipartite network that ignores the V-motifs constituted by, at least, one missing tie - hence, considering the bipartite topology as fixed, solely accounting for the signs of the `full' dyadic motifs. In the present section, differently from the main text, we will examine in detail all different cases, i.e. either when the statistical benchmark used employs global constraints or when it employs local constraints. 

\subsection{Benchmarks with global constraints}

A benchmark with global constraints is defined by the finite scheme reading

\begin{equation}
b_{i\alpha}\sim
\begin{pmatrix}
-1  & +1\\
1-p^+ & p^+
\end{pmatrix},\quad\forall\:i,\alpha\text{\:\:s.t.\:\:}|b_{i\alpha}|=1
\end{equation}
and further inducing

\begin{align}
b_{i\alpha}b_{j\alpha}\sim
\begin{pmatrix}
-1 & +1\\
2p^+(1-p^+) & 1-2p^++2(p^+)^2
\end{pmatrix}=\begin{pmatrix}
-1 & +1\\
1-q^+ & q^+
\end{pmatrix},\quad\forall\:i,j,\alpha\text{\:\:s.t.\:\:}|b_{i\alpha}b_{j\alpha}|=1;
\end{align}
naturally, $\langle b_{i\alpha}b_{j\alpha}\rangle=2q^+-1$ and $\text{Var}[b_{i\alpha}b_{j\alpha}]=4q^+(1-q^+)$.\\

Let us, now, notice that $\underline{S_{ij}}$ is a sum of i.i.d. Bernoulli random variables, the outcomes of each elementary event being, in fact, $-1$ and $+1$ instead of $0$ and $+1$: such a consideration allows us to find the probability distribution obeyed by $\underline{S_{ij}}$ quite straightforwardly, by starting from the one describing $\underline{C_{ij}}$ alone, i.e.

\begin{align}
P(\underline{C_{ij}}=k)=\binom{V_{ij}}{k}(q^+)^k(1-q^+)^{V_{ij}-k},
\end{align}
and considering the change of variable $k\equiv(V_{ij}+s)/2$, this, in turn, leads to the binomial

\begin{align}
P(\underline{S_{ij}}=s)=\binom{V_{ij}}{\frac{V_{ij}+s}{2}}(q^+)^\frac{V_{ij}+s}{2}(1-q^+)^\frac{V_{ij}-s}{2}
\end{align}
with $s$ ranging from $-V_{ij}$ to $V_{ij}$. As we said, in case $V_{ij}$ is odd, $s$ itself has to be odd, assuming the values $s=-V_{ij}$, $s=-V_{ij}+2$, $s=-V_{ij}+4$\dots up to $s=V_{ij}$\footnote{To convince ourselves that this is the case, let us, first, pose $V_{ij}=11$. Upon indicating with $c$ the number of concordant motifs and with $d$ the number of discordant motifs, the list of possible outcomes is, then, $(c,d)=(0,11)$ and $s=-11$; $(c,d)=(1,10)$ and $s=-9$; $(c,d)=(2,9)$ and $s=-7$; $(c,d)=(3,8)$ and $s=-5$; $(c,d)=(4,7)$ and $s=-3$; $(c,d)=(5,6)$ and $s=-1$; $(c,d)=(6,5)$ and $s=+1$; $(c,d)=(7,4)$ and $s=+3$; $(c,d)=(8,3)$ and $s=+5$; $(c,d)=(9,2)$ and $s=+7$; $(c,d)=(10,1)$ and $s=+9$; $(c,d)=(11,0)$ and $s=+11$.}. In words, differences are odd and are separated by steps of length two. In case $V_{ij}$ is even, $s$ itself has to be even, assuming the same values as above in correspondence of the same values of $k$\footnote{Again, to convince ourselves that this is the case, let us pose $V_{ij}=12$. The list of possible outcomes is, then, $(c,d)=(0,12)$ and $s=-12$; $(c,d)=(1,11)$ and $s=-10$; $(c,d)=(2,10)$ and $s=-8$; $(c,d)=(3,9)$ and $s=-6$; $(c,d)=(4,8)$ and $s=-4$; $(c,d)=(5,7)$ and $s=-2$; $(c,d)=(6,6)$ and $s=0$; $(c,d)=(7,5)$ and $s=+2$; $(c,d)=(8,4)$ and $s=+4$; $(c,d)=(9,3)$ and $s=+6$; $(c,d)=(10,2)$ and $s=+8$; $(c,d)=(11,1)$ and $s=+10$; $(c,d)=(12,0)$ and $s=+12$.}. In words, differences are even and are separated by steps of length two.\\

The binomial above can be approximated by a Gaussian whose parameters read

\begin{align}
\langle\underline{S_{ij}}\rangle=\sum_{\underline{\alpha=1}}^{V_{ij}}\langle b_{i\alpha}b_{j\alpha}\rangle=\sum_{\underline{\alpha=1}}^{V_{ij}}(2q^+-1)=V_{ij}(2q^+-1),
\end{align}
as further confirmed by posing $k\equiv(V_{ij}+s)/2$ and solving

\begin{align}\label{muGaussianHomo}
\langle\underline{S_{ij}}\rangle&=\sum_{s=-V_{ij}}^{V_{ij}}s\cdot\binom{V_{ij}}{\frac{V_{ij}+s}{2}}(q^+)^\frac{V_{ij}+s}{2}(q^-)^\frac{V_{ij}-s}{2}\nonumber\\
&=\sum_{k=0}^{V_{ij}}(2k-V_{ij})\cdot\binom{V_{ij}}{k}(q^+)^k(1-q^+)^{V_{ij}-k}\nonumber\\
&=V_{ij}(2q^+-1),\\
\end{align}
and

\begin{align}\label{varGaussianHomo}
\text{Var}[S_{ij}]&=\sum_{\underline{\alpha=1}}^V\text{Var}[b_{i\alpha}b_{j\alpha}]=\sum_{\underline{\alpha=1}}^{V_{ij}}4q^+(1-q^+)=4V_{ij}q^+(1-q^+),
\end{align}
as further confirmed by posing $k\equiv(V_{ij}+s)/2$ and solving

\begin{align}
\langle(\underline{S_{ij}})^2\rangle&=\sum_{s=-V_{ij}}^{V_{ij}}s^2\cdot\binom{V_{ij}}{\frac{V_{ij}+s}{2}}(q^+)^\frac{V_{ij}+s}{2}(1-q^+)^\frac{V_{ij}-s}{2}\nonumber\\
&=\sum_{k=0}^{V_{ij}}(2k-V_{ij})^2\cdot\binom{V_{ij}}{k}(q^+)^k(1-q^+)^{V_{ij}-k}\nonumber\\
&=4V_{ij}q^+-4V_{ij}(q^+)^2+4V_{ij}^2(q^+)^2-4V_{ij}^2q^++V^2;
\end{align}
subtracting $\langle\underline{S_{ij}}\rangle^2$ from it leads to recover the expression $\text{Var}[S_{ij}]=4V_{ij}q^+(1-q^+)$ (see Fig.~\ref{fig_A_1}).

\subsection{Benchmarks with local constraints}

A benchmark with local constraints is defined by the finite scheme reading

\begin{equation}
b_{i\alpha}\sim
\begin{pmatrix}
-1  & +1\\
1-p_{i\alpha}^+ & p_{i\alpha}^+
\end{pmatrix},\quad\forall\:i,\alpha\text{\:\:s.t.\:\:}|b_{i\alpha}|=1,
\end{equation}
further inducing

\begin{align}
b_{i\alpha}b_{j\alpha}\sim
\begin{pmatrix}
-1 & +1\\
p_{i\alpha}^++p_{j\alpha}^+-2p_{i\alpha}^+p_{j\alpha}^+ & 1-p_{i\alpha}^+-p_{j\alpha}^++2p_{i\alpha}^+p_{j\alpha}^+\end{pmatrix}=
\begin{pmatrix}
-1 & +1\\
1-q_{ij\alpha}^+ & q_{ij\alpha}^+
\end{pmatrix},\quad\forall\:i,j,\alpha\text{\:\:s.t.\:\:}|b_{i\alpha}b_{j\alpha}|=1;
\end{align}
naturally, $\langle b_{i\alpha}b_{j\alpha}\rangle=2q_{ij\alpha}^+-1$ and $\text{Var}[b_{i\alpha}b_{j\alpha}]=4q_{ij\alpha}^+(1-q_{ij\alpha}^+)$.\\

Since $\underline{S_{ij}}$ is a sum of i.n.i.d. Bernoulli random variables, the outcomes of each, elementary event being $-1$ and $+1$, the probability distribution obeyed by $\underline{S_{ij}}$ is the Poisson-binomial

\begin{align}
P(\underline{S_{ij}}=s)=\sum_{C_k\in\mathcal{C}_k}\left[\prod_{\nu\in C_k}q_{ij\nu}^+\prod_{\tau\notin C_k}(1-q_{ij\tau}^+)\right]
\end{align}
where $\mathcal{C}_k$ is the set of all possible $k$-tuples of which $\nu$ and $\tau$ are instances. Let us provide some explicit examples:

\begin{align*}
&P(\underline{S_{ij}}=-V_{ij})=\prod_{\underline{\alpha=1}}^{V_{ij}}(1-q_{ij\alpha}^+),\\
&P(\underline{S_{ij}}=2-V_{ij})=\sum_{\underline{\beta=1}}^{V_{ij}}\left[q_{ij\beta}^+\prod_{\substack{\underline{\alpha=1}\\\alpha\neq\beta}}^{V_{ij}}(1-q_{ij\alpha}^+)\right],\\
&P(\underline{S_{ij}}=4-V_{ij})=\sum_{\underline{\beta=1}}^{V_{ij}}\sum_{\substack{\underline{\gamma=1}\\\gamma>\beta}}^{V_{ij}}\left[q_{ij\beta}^+q_{ij\gamma}^+\prod_{\substack{\underline{\alpha=1}\\\alpha\neq\beta,\gamma}}^{V_{ij}}(1-q_{ij\alpha}^+)\right]
\end{align*}
and so on. In words, if nodes $i$ and $j$ establish $k=0$ `full' concordant motifs, they establish $V_{ij}$ `full' discordant motifs and the signature reads $s=-V_{ij}$; if nodes $i$ and $j$ establish $k=1$ `full' concordant motif, they establish $V_{ij}-1$ `full' discordant motifs and the signature reads $s=2-V_{ij}$; if nodes $i$ and $j$ establish $k=2$ `full' concordant motif, they establish $V_{ij}-2$ `full' discordant motifs and the signature reads $s=4-V_{ij}$. Analogously for the other admissible values of $k$.\\

\begin{figure*}[t!]
\centering
\includegraphics[width=\textwidth]{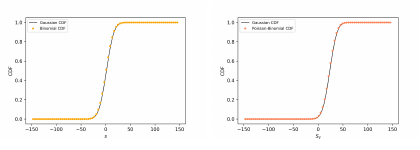}
\caption{Left panel: comparison between the CDF of the binomial distribution obeyed by $S_{ij}$ under the global statistical benchmark induced by the zero-deflated scheme and its Gaussian approximation. Right panel: comparison between the CDF of the Poisson-binomial distribution obeyed by $S_{ij}$ under the local statistical benchmark induced by the zero-deflated scheme and its Gaussian approximation. In both cases, the distribution is defined by the parameters induced by nodes $i=3$ and $j=31$ of the \textit{U.S. Senate} dataset ($V_{ij}=147$).}
\label{fig_A_1}
\end{figure*}

Such a distribution can be approximated by a Gaussian whose moments read

\begin{align}\label{muGaussianHete}
\langle\underline{S_{ij}}\rangle=\sum_{\underline{\alpha=1}}^{V_{ij}}\langle b_{i\alpha}b_{j\alpha}\rangle=\sum_{\underline{\alpha=1}}^{V_{ij}}(2q_{ij\alpha}^+-1),\\
\label{varGaussianHete}
\text{Var}[\underline{S_{ij}}]=\sum_{\underline{\alpha=1}}^{V_{ij}}\text{Var}[b_{i\alpha}b_{j\alpha}]=\sum_{\underline{\alpha=1}}^{V_{ij}}4q_{ij\alpha}^+(1-q_{ij\alpha}^+)
\end{align}
(see Fig.~\ref{fig_A_1}).

\clearpage

\section{The distribution of the signature in the zero-inflated scheme}\label{AppB}

Let us, first, remind that such a scheme devises a recipe for projecting a bipartite network that accounts for the V-motifs constituted by, at least, one missing tie as well - hence, considering the bipartite topology as free, accounting for the signs of both the `full' and the `partial' dyadic motifs. As in the previous section, we will examine in detail all different cases, i.e. either when the statistical benchmark used employs global constraints or when it employs local constraints. 

\subsection{Benchmarks with global constraints}

A benchmark with global constraints is defined by the finite scheme reading

\begin{equation}
b_{i\alpha}\sim
\begin{pmatrix}
-1 & 0 & +1\\
p^- & p^0 & p^+
\end{pmatrix},\quad\forall\:i,\alpha
\end{equation}
that, in turn, leads to

\begin{align}
b_{i\alpha}^+b_{j\alpha}^+&\sim
\begin{pmatrix}
0 & +1\\
1-(p^+)^2 & (p^+)^2
\end{pmatrix}=\begin{pmatrix}
0 & +1\\
1-q^{++} & q^{++}
\end{pmatrix},\quad\forall\:i,j,\alpha,\\
b_{i\alpha}^-b_{j\alpha}^-&\sim
\begin{pmatrix}
0 & +1\\
1-(p^-)^2 & (p^-)^2
\end{pmatrix}=\begin{pmatrix}
0 & +1\\
1-q^{--} & q^{--}
\end{pmatrix},\quad\forall\:i,j,\alpha,\\
b_{i\alpha}^0b_{j\alpha}^0&\sim
\begin{pmatrix}
0 & +1\\
1-(p^0)^2 & (p^0)^2
\end{pmatrix}=\begin{pmatrix}
0 & +1\\
1-q^{00} & q^{00}
\end{pmatrix},\quad\forall\:i,j,\alpha
\end{align}
for what concerns concordant motifs and 

\begin{align}
b_{i\alpha}^+b_{j\alpha}^-&\sim
\begin{pmatrix}
0 & +1\\
1-p^+p^- & p^+p^-
\end{pmatrix}=\begin{pmatrix}
0 & +1\\
1-q^{+-} & q^{+-}
\end{pmatrix},\quad\forall\:i,j,\alpha,\\
b_{i\alpha}^0b_{j\alpha}^+&\sim
\begin{pmatrix}
0 & +1\\
1-p^0p^+ & p^0p^+
\end{pmatrix}=\begin{pmatrix}
0 & +1\\
1-q^{0+} & q^{0+}
\end{pmatrix},\quad\forall\:i,j,\alpha,\\
b_{i\alpha}^0b_{j\alpha}^-&\sim
\begin{pmatrix}
0 & +1\\
1-p^0p^- & p^0p^-
\end{pmatrix}=\begin{pmatrix}
0 & +1\\
1-q^{0-} & q^{0-}
\end{pmatrix},\quad\forall\:i,j,\alpha
\end{align}
for what concerns discordant motifs. Naturally, $\langle b_{i\alpha}^+b_{j\alpha}^+\rangle=q^{++}$ and $\text{Var}[b_{i\alpha}^+b_{j\alpha}^+]=q^{++}(1-q^{++})$ and analogously for the other motifs. One may, now, define the finite schemes obeyed by the triple-wise concordance and discordance as

\begin{align}
C_{ij\alpha}&=b_{i\alpha}^+b_{j\alpha}^++b_{i\alpha}^-b_{j\alpha}^-+b_{i\alpha}^0b_{j\alpha}^0\sim
\begin{pmatrix}
0 & +1\\
1-q^+ & q^+
\end{pmatrix},\quad\forall\:i,j,\alpha,\\
D_{ij\alpha}&=b_{i\alpha}^+b_{j\alpha}^-+b_{i\alpha}^-b_{j\alpha}^++b_{i\alpha}^0b_{j\alpha}^++b_{i\alpha}^+b_{j\alpha}^0+b_{i\alpha}^0b_{j\alpha}^-+b_{i\alpha}^-b_{j\alpha}^0\sim\begin{pmatrix}
0 & +1\\
1-q^- & q^-
\end{pmatrix},\quad\forall\:i,j,\alpha
\end{align}
the coefficients therein reading

\begin{equation}
q^-=q^{+-}+q^{-+}+q^{0+}+q^{+0}+q^{0-}+q^{-0}=2p^+p^-+2p^0p^++2p^0p^-
\end{equation}
and

\begin{equation}
q^+=q^{++}+q^{--}+q^{00}=(p^+)^2+(p^-)^2+(p^0)^2;
\end{equation}
naturally, $q^-+q^+=(p^-+p^0+p^+)^2=1$. As a consequence of such definitions, $S_{ij}=\sum_{\alpha=1}^M(C_{ij\alpha}-D_{ij\alpha})=C_{ij}-D_{ij}$ obeys the distribution

\begin{align}
P(S_{ij}=s)=\binom{M}{\frac{M+s}{2}}(q^+)^\frac{M+s}{2}(q^-)^\frac{M-s}{2}
\end{align}
whose moments read $\langle S_{ij}\rangle=M(q^+-q^-)$ and $\text{Var}[S_{ij}]=4Mq^+q^-$.\\

Naturally, the Gaussian approximation of the binomial above is defined by the moments

\begin{align}
\langle S_{ij}\rangle=\sum_{\alpha=1}^M\langle C_{ij\alpha}-D_{ij\alpha}\rangle=M(q^+-q^-),\\
\text{Var}[S_{ij}]=\sum_{\alpha=1}^M\text{Var}[C_{ij\alpha}-D_{ij\alpha}]=4Mq^+q^-
\end{align}
(see Fig.~\ref{fig_A_2}).

\subsection{Benchmarks with local constraints}

A benchmark with local constraints is defined by the finite scheme reading

\begin{align}
C_{ij\alpha}&=b_{i\alpha}^+b_{j\alpha}^++b_{i\alpha}^-b_{j\alpha}^-+b_{i\alpha}^0b_{j\alpha}^0\sim
\begin{pmatrix}
0 & +1\\
1-q_{ij\alpha}^+ & q_{ij\alpha}^+
\end{pmatrix},\quad\forall\:i,j,\alpha,\\
D_{ij\alpha}&=b_{i\alpha}^+b_{j\alpha}^-+b_{i\alpha}^-b_{j\alpha}^++b_{i\alpha}^0b_{j\alpha}^++b_{i\alpha}^+b_{j\alpha}^0+b_{i\alpha}^0b_{j\alpha}^-+b_{i\alpha}^-b_{j\alpha}^0\sim\begin{pmatrix}
0 & +1\\
1-q_{ij\alpha}^- & q_{ij\alpha}^-
\end{pmatrix},\quad\forall\:i,j,\alpha
\end{align}
the coefficients therein reading

\begin{equation}
q_{ij\alpha}^-=q_{ij\alpha}^{+-}+q_{ij\alpha}^{-+}+q_{ij\alpha}^{0+}+q_{ij\alpha}^{+0}+q_{ij\alpha}^{0-}+q_{ij\alpha}^{-0}=p_{i\alpha}^+p_{j\alpha}^-+p_{i\alpha}^-p_{j\alpha}^++p_{i\alpha}^0p_{j\alpha}^++p_{i\alpha}^+p_{j\alpha}^0+p_{i\alpha}^0p_{j\alpha}^-+p_{i\alpha}^-p_{j\alpha}^0
\end{equation}
and

\begin{equation}
q_{ij\alpha}^+=q_{ij\alpha}^{++}+q_{ij\alpha}^{--}+q_{ij\alpha}^{00}=p_{i\alpha}^+p_{j\alpha}^++p_{i\alpha}^-p_{j\alpha}^-+p_{i\alpha}^0p_{j\alpha}^0;
\end{equation}
naturally,

\begin{align}
q_{ij\alpha}^-+q_{ij\alpha}^+&=p_{i\alpha}^+(p_{j\alpha}^++p_{j\alpha}^0+p_{j\alpha}^-)+p_{i\alpha}^0(p_{j\alpha}^++p_{j\alpha}^0+p_{j\alpha}^-)+p_{i\alpha}^-(p_{j\alpha}^++p_{j\alpha}^0+p_{j\alpha}^-)\nonumber\\
&=(p_{i\alpha}^++p_{i\alpha}^0+p_{i\alpha}^-)(p_{j\alpha}^++p_{j\alpha}^0+p_{j\alpha}^-)\nonumber\\
&=1\cdot1\nonumber\\
&=1.
\end{align}

As a consequence of such definitions, $S_{ij}=\sum_{\alpha=1}^M(C_{ij\alpha}-D_{ij\alpha})=C_{ij}-D_{ij}$ obeys the distribution

\begin{align}
P(S_{ij}=s)=\sum_{C_k\in\mathcal{C}_k}\left[\prod_{\nu\in C_k}q_{ij\nu}^+\prod_{\tau\notin C_k}q_{ij\tau}^-\right]
\end{align}
where $\mathcal{C}_k$ is the set of all possible $k$-tuples of which $\nu$ and $\tau$ are instances. More explicitly,

\begin{align}
&P(S_{ij}=-M)=\prod_{\alpha=1}^Mq_{ij\alpha}^-,\\
&P(S_{ij}=2-M)=\sum_{\beta=1}^M\left[q_{ij\beta}^+\prod_{\substack{\alpha=1\\\alpha\neq\beta}}^Mq_{ij\alpha}^-\right],\\
&P(S_{ij}=4-M)=\sum_{\beta=1}^M\sum_{\substack{\gamma=1\\\gamma>\beta}}^M\left[q_{ij\beta}^+q_{ij\gamma}^+\prod_{\substack{\alpha=1\\\alpha\neq\beta,\gamma}}^Mq_{ij\alpha}^-\right]
\end{align}
and so on. In words, if nodes $i$ and $j$ establish $k=0$ concordant motifs, they establish $M$ discordant motifs and the signature reads $s=-M$; if nodes $i$ and $j$ establish $k=1$ concordant motif, they establish $M-1$ discordant motifs and the signature reads $s=2-M$; if nodes $i$ and $j$ establish $k=2$ `full' concordant motif, they establish $M-2$ `full' discordant motifs and the signature reads $s=4-M$. The same line of reasoning can be repeated for all the admissible values of $k$.\\

\begin{figure*}[t!]
\centering
\includegraphics[width=\textwidth]{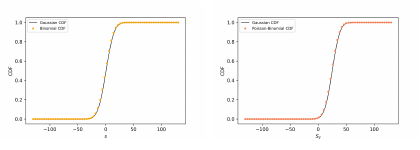}
\caption{Left panel: comparison between the CDF of the binomial distribution obeyed by $S_{ij}$ under the global statistical benchmark induced by the zero-inflated scheme and its Gaussian approximation. Right panel: comparison between the CDF of the Poisson-binomial distribution obeyed by $S_{ij}$ under the local statistical benchmark induced by the zero-inflated scheme and its Gaussian approximation. In both cases, the distribution is defined by the parameters induced by nodes $i=11$ and $j=16$ of the \textit{U.S. House of Representatives} dataset ($V_{ij}=130$).}
\label{fig_A_2}
\end{figure*}

Such a distribution can be approximated by a Gaussian whose moments read

\begin{align}
\langle S_{ij}\rangle=\sum_{\alpha=1}^M\langle C_{ij\alpha}-D_{ij\alpha}\rangle=\sum_{\alpha=1}^M(q_{ij\alpha}^+-q_{ij\alpha}^-),\\
\text{Var}[S_{ij}]=\sum_{\alpha=1}^M\text{Var}[C_{ij\alpha}-D_{ij\alpha}]=\sum_{\alpha=1}^M4q_{ij\alpha}^+q_{ij\alpha}^-
\end{align}
since

\begin{align}
\text{Var}[C_{ij\alpha}-D_{ij\alpha}]&=\text{Var}[C_{ij\alpha}]+\text{Var}[D_{ij\alpha}]-2\cdot\text{Cov}[C_{ij\alpha},D_{ij\alpha}]\nonumber\\
&=\text{Var}[C_{ij\alpha}]+\text{Var}[D_{ij\alpha}]-2\cdot[\langle C_{ij}\cdot D_{ij}\rangle-\langle C_{ij}\rangle\langle D_{ij}\rangle]\nonumber\\
&=\text{Var}[C_{ij\alpha}]+\text{Var}[D_{ij\alpha}]-2\cdot\langle C_{ij}\rangle\langle D_{ij}\rangle\nonumber\\
&=q_{ij\alpha}^+q_{ij\alpha}^-+q_{ij\alpha}^+q_{ij\alpha}^-+2q_{ij\alpha}^+q_{ij\alpha}^-\nonumber\\
&=4q_{ij\alpha}^+q_{ij\alpha}^-
\end{align}
(see Fig.~\ref{fig_A_2}).

\clearpage

\section{Multiple hypothesis testing: the False Discovery Rate}\label{App:FDR}

Whenever $|H|$ different hypotheses $H_1,H_2\dots$, characterised by $|H|$ different $p-$values, have to be tested at the same time, FDR~\cite{benjamini1995controlling} prescribes to, first, sort the $|H|$ $p-$values in increasing order

\begin{align}
\text{$p-$value}_1\le\dots\le\text{$p-$value}_{|H|}
\end{align}
and, then, identify the largest integer $\hat{i}$ satisfying the condition

\begin{align}
\text{$p-$value}_{\hat{i}}\le\frac{\hat{i}t}{|H|}
\end{align}
with $t$ representing the usual single-test significance level (hereby set to the value $t=0.05$). Since $H_i$ is the hypothesis that the distribution of dyadic motifs established by the $i$th pair of nodes follows a binomial/Poisson-binomial, rejecting it amounts at connecting the corresponding nodes in the projection~\cite{saracco2017inferring}.

Notice that $|H|=\sum_{i=1}^N\sum_{j(>i)}V_{ij}$ within the zero-deflated projection scheme and $|H|=N(N-1)/2$ within the zero-inflated projection scheme.

\clearpage

\section{Probabilistic models for binary undirected bipartite signed networks}\label{AppC}

The generalisation of the ERG formalism for the analysis of bipartite signed graphs rests upon the constrained maximisation of Shannon entropy

\begin{equation}
\mathscr{L}=S[P]-\sum_{i=0}^T\theta_i[P(\mathbf{B})C_i(\mathbf{B})-\langle C_i\rangle]
\end{equation}
where $S=-\sum_{\mathbf{B}\in\mathbb{B}}P(\mathbf{B})\ln P(\mathbf{B})$, $C_0=\langle C_0\rangle=1$ sums up the normalisation condition and the remaining $T-1$ constraints represent proper topological properties. Such an optimisation procedure defines the expression

\begin{equation}
P(\mathbf{B})=\frac{e^{-H(\bm{\theta},\mathbf{B})}}{Z(\bm{\theta})}=\frac{e^{-H(\bm{\theta},\mathbf{B})}}{\sum_{\mathbf{B}\in\mathbb{B}}e^{-H(\bm{\theta},\mathbf{B})}}=\frac{e^{-\sum_{i=1}^T\theta_iC_i(\mathbf{B})}}{\sum_{\mathbf{B}\in\mathbb{B}}e^{-\sum_{i=1}^T\theta_iC_i(\mathbf{B})}}
\end{equation}
that can be made explicit only after a specific set of constraints has been chosen.

\subsection{Free-topology Bipartite Signed Random Graph Model (BiSRGM)}

The Hamiltonian describing such a problem reads

\begin{equation}
H(\bm{\theta},\mathbf{B})=\beta L^+(\mathbf{B})+\gamma L^-(\mathbf{B})=\sum_{i=1}^N\sum_{\alpha=1}^M(\beta b_{i\alpha}^++\gamma b_{i\alpha}^-);
\end{equation}
as a consequence, the partition function reads

\begin{align}
Z(\bm{\theta})&=\sum_{\mathbf{B}\in\mathbb{B}}\prod_{i=1}^N\prod_{\alpha=1}^M e^{-(\beta b_{i\alpha}^++\gamma b_{i\alpha}^-)}=\prod_{i=1}^N\prod_{\alpha=1}^M\sum_{b_{i\alpha}=-1,0,1}e^{-(\beta b_{i\alpha}^++\gamma b_{i\alpha}^-)}=\prod_{i=1}^N\prod_{\alpha=1}^M(1+e^{-\beta}+e^{-\gamma})=(1+e^{-\beta}+e^{-\gamma})^{N\cdot M}
\end{align}
and induces the expression

\begin{equation}
P_\text{BiSRGM}(\mathbf{B})=\frac{e^{-\beta L^+(\mathbf{B})-\gamma L^-(\mathbf{B})}}{(1+e^{-\beta}+e^{-\gamma})^{N\cdot M}}=\frac{x^{L^+(\mathbf{B})}y^{L^-(\mathbf{B})}}{(1+x+y)^{N\cdot M}}=(p^-)^{L^-}(p^0)^{L^0}(p^+)^{L^+}
\end{equation}
having posed

\begin{align}
p^-&=\frac{e^{-\gamma}}{1+e^{-\beta}+e^{-\gamma}}=\frac{y}{1+x+y},\\
p^0&=\frac{1}{1+e^{-\beta}+e^{-\gamma}}=\frac{1}{1+x+y},\\
p^+&=\frac{e^{-\beta}}{1+e^{-\beta}+e^{-\gamma}}=\frac{x}{1+x+y}
\end{align}
where $p^+$ is the probability that a node belonging to the first set and a node belonging to the second set are linked by a positive edge, $p^-$ is the probability that a node belonging to the first set and a node belonging to the second set are linked by a negative edge and $p^0$ is the probability that a node belonging to the first set and a node belonging to the second set are no linked at all. Naturally, $p^0=1-p^--p^+$. Hence, according to the BiSRGM, each entry of a bipartite signed network is a random variable following a generalised Bernoulli distribution, i.e. obeying the finite scheme

\begin{equation}
b_{i\alpha}\sim
\begin{pmatrix}
-1 & 0 & +1\\
p^- & p^0 & p^+
\end{pmatrix},\quad\forall\:i,\alpha.
\end{equation}

To determine the parameters that define the BiSRGM, let us maximise the likelihood function 

\begin{align}
\mathcal{L}_\text{BiSRGM}(x,y)=\ln P_\text{BiSRGM}(\mathbf{B}^*|x,y)=L^+(\mathbf{B}^*)\ln(x)+L^-(\mathbf{B}^*)\ln(y)-(N\cdot M)\ln(1+x+y)
\end{align}
with respect to $x$ and $y$. Upon doing so, we obtain the pair of equations

\begin{align}
\frac{\partial\mathcal{L}_\text{BiSRGM}(x,y)}{\partial x}&=\frac{L^+(\mathbf{B}^*)}{x}-(N\cdot M)\frac{1}{1+x+y},\\
\frac{\partial\mathcal{L}_\text{BiSRGM}(x,y)}{\partial y}&=\frac{L^-(\mathbf{B}^*)}{y}-(N\cdot M)\frac{1}{1+x+y};
\end{align}
equating them to zero leads us to find $L^+(\mathbf{B}^*)=(N\cdot M)\frac{x}{1+x+y}=(N\cdot M)p^+=\langle L^+\rangle$ and $L^-(\mathbf{B}^*)=(N\cdot M)\frac{y}{1+x+y}=(N\cdot M)p^-=\langle L^-\rangle$, i.e. 

\begin{align}
p^+&=L^+(\mathbf{B}^*)/(N\cdot M),\\
p^-&=L^-(\mathbf{B}^*)/(N\cdot M).
\end{align}

\subsection{Fixed-topology Bipartite Signed Random Graph Model (BiSRGM-FT)}

Let us, again, consider the properties $L^+(\mathbf{B})$ and $L^-(\mathbf{B})$, to be satisfied by keeping a bipartite network topology fixed. The Hamiltonian describing such a problem still reads

\begin{equation}\label{eq:H_BISRGM_FT}
H(\bm{\theta},\mathbf{B})=\beta' L^+(\mathbf{B})+\gamma' L^-(\mathbf{B})=\sum_{i=1}^N\sum_{\alpha=1}^M(\beta' b_{i\alpha}^++\gamma' b_{i\alpha}^-).
\end{equation}

Since the topology is fixed, we can solely focus on the non-zero entries of the biadjacency matrix $\mathbf{B}$: this choice allows us to simplify the Hamiltonian in \ref{eq:H_BISRGM_FT} by posing

\begin{equation}
H(\bm{\theta},\mathbf{B})=\sum_{i=1}^N\sum_{\alpha=1}^M(\beta'b_{i\alpha}^++\gamma'b_{i\alpha}^-)=\sum_{i=1}^N\sum_{\alpha=1}^M\left[\beta'b_{i\alpha}^++\gamma'(1-b_{i\alpha}^+)\right];
\end{equation}
dropping the constant term, playing no role in determining the probabilities, and defining $\beta=\beta'-\gamma'$ leads to the partition function reading

\begin{align}
Z(\bm{\theta})=&\sum_{\substack{\mathbf{B}\in\mathbb{B}\\(|\mathbf B|=|\mathbf B^*|)}}\prod_{i=1}^N\prod_{\alpha=1}^Me^{-\beta b_{i\alpha}^+}=\prod_{i=1}^N\prod_{\alpha=1}^N\left(\sum_{b_{i\alpha}=-1,1}e^{-\beta b_{i\alpha}^+}\right)^{|b_{i\alpha}^*|}=\prod_{i=1}^N\prod_{\alpha=1}^M(1+e^{-\beta})^{|b_{i\alpha}^*|}=(1+e^{-\beta})^L
\end{align}
and to the expression

\begin{align}\label{probgraph}
P_\text{BiSRGM-FT}(\mathbf{B})=\frac{e^{-\beta L^+(\mathbf{B})}}{(1+e^{-\beta})^L}=\frac{x^{L^+(\mathbf{B})}}{(1+x)^L}=(1-p^+)^{L-L^+}(p^+)^{L^+},
\end{align}
having posed $x=e^{-\beta}$ and 

\begin{equation*}
p^+=\frac{e^{-\beta}}{1+e^{-\beta}}=\frac{x}{1+x}
\end{equation*} 
(the probability that any two connected nodes are linked by a negative edge is $p^-=1-p^+$). Hence, according to the BiSRGM-FT, each entry of a bipartite signed network such that $|b_{i\alpha}^*|=1$ is a random variable following a Bernoulli distribution that obeys the finite scheme

\begin{equation}
b_{i\alpha}\sim
\begin{pmatrix}
-1 & +1\\
p^- & p^+
\end{pmatrix},\quad\forall\:i,\alpha\text{\:\:s.t.\:\:}|b_{i\alpha}^*|=1.
\end{equation}

In order to determine the parameters that define the BiSRGM-FT, let us maximise the likelihood function

\begin{equation}
\mathcal{L}_\text{BiSRGM-FT}(x)=\ln P_\text{BiSRGM-FT}(\mathbf{B}^*|x)=L^+(\mathbf{B}^*)\ln(x)+-L(\mathbf{B}^*)\ln(1+x)
\end{equation}
with respect to $x$. Upon doing so, we obtain 

\begin{equation*}
\frac{\partial\mathcal{L}_\text{BiSRGM-FT}(x)}{\partial x}=\frac{L^+(\mathbf{B}^*)}{x}-\frac{L(\mathbf{B}^*)}{1+x}.
\end{equation*}
Equating it to zero leads us to find $L^+(\mathbf{B}^*)=L(\mathbf{B}^*)\frac{x}{x+y}=L(\mathbf{B}^*)p^+=\langle L^+\rangle$, i.e. 

\begin{equation*}
p^+=L^+(\mathbf{B}^*)/L(\mathbf{B}^*).
\end{equation*}

\subsection{Free-topology Bipartite Signed Configuration Model (BiSCM)}

The second set of constraints we consider is represented by the properties $\{k_i^+(\mathbf{B})\}_{i=1}^N$, $\{k_i^-(\mathbf{B})\}_{i=1}^N$, $\{h_\alpha^+(\mathbf{B})\}_{\alpha=1}^M$ and $\{h_\alpha^-(\mathbf{B})\}_{\alpha=1}^M$. The Hamiltonian describing such a problem reads

\begin{align}
H(\bm{\theta},\mathbf{B})=\sum_{i=1}^N[\beta_ik_i^+(\mathbf B)+\gamma_ik_i^-(\mathbf B)]+\sum_{\alpha=1}^M[\delta_\alpha h_\alpha^+(\mathbf B)+\eta_\alpha h_\alpha^-(\mathbf B)].
\end{align}
As a consequence, the partition function reads

\begin{align}
Z(\bm{\theta})&=\sum_{\mathbf{B}\in\mathbb B}\prod_{i=1}^N\prod_{\alpha=1}^Me^{-(\beta_i+\delta_\alpha)b_{i\alpha}^+-(\gamma_i+\eta_\alpha)b_{i\alpha}^-}=\prod_{i=1}^N\prod_{\alpha=1}^M\sum_{b_{i\alpha}=-1,0,1}e^{-(\beta_i+\delta_\alpha)b_{i\alpha}^+-(\gamma_i+\eta_\alpha)b_{i\alpha}^-}=\prod_{i=1}^N\prod_{\alpha=1}^M\left[1+e^{-(\beta_i+\delta_\alpha)}+e^{-(\gamma_i+\eta_\alpha)}\right]
\end{align}
and induces the expression

\begin{equation*}
P_\text{BiSCM}(\mathbf{B})=\prod_{i=1}^N\prod_{\alpha=1}^M(p_{i\alpha}^-)^{b_{i\alpha}^-}(p_{i\alpha}^0)^{b_{i\alpha}^0}(p_{i\alpha}^+)^{b_{i\alpha}^+}
\end{equation*}
having posed

\begin{align}
p_{i\alpha}^-&=\frac{e^{-(\gamma_i+\eta_\alpha)}}{1+e^{-(\beta_i+\delta_\alpha)}+e^{-(\gamma_i+\eta_\alpha)}}=\frac{y_iw_\alpha}{1+x_iz_\alpha+y_iw_\alpha},\\
p_{i\alpha}^0&=\frac{1}{1+e^{-(\beta_i+\delta_\alpha)}+e^{-(\gamma_i+\eta_\alpha)}}=\frac{1}{1+x_iz_\alpha+y_iw_\alpha},\\
p_{i\alpha}^+&=\frac{e^{-(\beta_i+\eta_\alpha)}}{1+e^{-(\beta_i+\delta_\alpha)}+e^{-(\gamma_i+\eta_\alpha)}}=\frac{x_iz_\alpha}{1+x_iz_\alpha+y_iw_\alpha}
\end{align}
where $p_{i\alpha}^+$ is the probability that nodes $i$ and $\alpha$ are linked by a positive edge, $p_{i\alpha}^-$ is the probability that nodes $i$ and $\alpha$ are linked by a negative edge and $p_{i\alpha}^0$ is the probability that nodes $i$ and $\alpha$ are no linked at all. Hence, according to the BiSCM, the generic entry of a bipartite signed network is a random variable following a generalised Bernoulli distribution, i.e. obeying the finite scheme

\begin{align}
b_{i\alpha}\sim
\begin{pmatrix}
-1 & 0 & +1\\
p_{i\alpha}^- & p_{i\alpha}^0 & p_{i\alpha}^+
\end{pmatrix},\quad\forall\:i,\alpha.
\end{align}

In order to determine the parameters that define the BiSCM, let us maximise the likelihood function

\begin{align}
&\mathcal{L}_\text{BiSCM}(\mathbf{x},\mathbf{y},\mathbf{z},\mathbf{w})=\ln P_\text{BiSCM}(\mathbf{x},\mathbf{y},\mathbf{z},\mathbf{w})\nonumber\\
&=\sum_{i=1}^Nk_i^+(\mathbf{B}^*)\ln(x_i)+\sum_{i=1}^Nk_i^-(\mathbf{B}^*)\ln(y_i)+\sum_{\alpha=1}^Mh_\alpha^+(\mathbf{B}^*)\ln(z_\alpha)+\sum_{\alpha=1}^Mh_\alpha^-(\mathbf{B}^*)\ln(w_\alpha)- \sum_{i=1}^N\sum_{\alpha=1}^M\ln(1+x_iz_\alpha+y_iw_\alpha)
\end{align}
with respect to $x_i$, $y_i$, $z_\alpha$ and $w_\alpha$, $\forall\:i,\alpha$. Upon doing so, we obtain the system of equations

\begin{align}
\frac{\partial\mathcal{L}_\text{BiSCM}(\{x_i\}_{i=1}^N,\{y_i\}_{i=1}^N,\{z_\alpha\}_{\alpha=1}^M,\{w_\alpha\}_{\alpha=1}^M)}{\partial x_i}&=\frac{k_i^+(\mathbf{B}^*)}{x_i}-\sum_{\alpha=1}^M\frac{z_\alpha}{1+x_iz_\alpha+y_iw_\alpha},\quad\forall\:i,\\
\frac{\partial\mathcal{L}_\text{BiSCM}(\{x_i\}_{i=1}^N,\{y_i\}_{i=1}^N,\{z_\alpha\}_{\alpha=1}^M,\{w_\alpha\}_{\alpha=1}^M)}{\partial y_i}&=\frac{k_i^-(\mathbf{B}^*)}{y_i}-\sum_{\alpha=1}^M\frac{w_\alpha}{1+x_iz_\alpha+y_iw_\alpha},\quad\forall\:i,\\
\frac{\partial\mathcal{L}_\text{BiSCM}(\{x_i\}_{i=1}^N,\{y_i\}_{i=1}^N,\{z_\alpha\}_{\alpha=1}^M,\{w_\alpha\}_{\alpha=1}^M)}{\partial z_\alpha}&=\frac{h_\alpha^+(\mathbf{B}^*)}{z_\alpha}-\sum_{i=1}^N\frac{x_i}{1+x_iz_\alpha+y_iw_\alpha},\quad\forall\:\alpha,\\
\frac{\partial\mathcal{L}_\text{BiSCM}(\{x_i\}_{i=1}^N,\{y_i\}_{i=1}^N,\{z_\alpha\}_{\alpha=1}^M,\{w_\alpha\}_{\alpha=1}^M)}{\partial w_\alpha}&=\frac{h_\alpha^-(\mathbf{B}^*)}{w_\alpha}-\sum_{i=1}^N\frac{y_i}{1+x_iz_\alpha+y_iw_\alpha},\quad\forall\:\alpha;
\end{align}
equating them to zero leads us to find

\begin{align}
k_i^+(\mathbf{B}^*)&=\sum_{\alpha=1}^M\frac{x_iz_\alpha}{1+x_iz_\alpha+y_iw_\alpha}=\sum_{\alpha=1}^Mp_{i\alpha}^+=\langle k_i^+\rangle,\quad\forall\:i,\\
k_i^-(\mathbf{B}^*)&=\sum_{\alpha=1}^M\frac{y_iw_\alpha}{1+x_iz_\alpha+y_iw_\alpha}=\sum_{\alpha=1}^Mp_{i\alpha}^-=\langle k_i^-\rangle,\quad\forall\:i,\\
h_\alpha^+(\mathbf{B}^*)&=\sum_{i=1}^N\frac{x_iz_\alpha}{1+x_iz_\alpha+y_iw_\alpha}=\sum_{i=1}^Np_{i\alpha}^+=\langle h_\alpha^+\rangle,\quad\forall\:\alpha,\\
h_\alpha^-(\mathbf{B}^*)&=\sum_{i=1}^N\frac{y_iw_\alpha}{1+x_iz_\alpha+y_iw_\alpha}=\sum_{i=1}^Np_{i\alpha}^-=\langle h_\alpha^-\rangle,\quad\forall\:\alpha.
\end{align}

\subsection{Fixed-topology Bipartite Signed Configuration Model (BiSCM-FT)}

Let us, again, consider the properties $\{k_i^+(\mathbf{B})\}_{i=1}^N$, $\{k_i^-(\mathbf{B})\}_{i=1}^N$, $\{h_\alpha^+(\mathbf{B})\}_{\alpha=1}^M$ and $\{h_\alpha^-(\mathbf{B})\}_{\alpha=1}^M$ to be satisfied by keeping a bipartite network topology fixed. The Hamiltonian describing such a problem still reads

\begin{align}
H(\bm{\theta},\mathbf{B})=\sum_{i=1}^N[\beta'_ik_i^+(\mathbf B)+\gamma'_ik_i^-(\mathbf B)]+\sum_{\alpha=1}^M[\delta'_\alpha h_\alpha^+(\mathbf B)+\eta'_\alpha h_\alpha^-(\mathbf B)], 
\end{align}
but can be simplified into

\begin{align}
H(\bm{\theta},\mathbf{B})=\sum_{i=1}^N\beta_ik_i^+(\mathbf B)+\sum_{\alpha=1}^M\delta_\alpha h_\alpha^+(\mathbf B);
\end{align}
as a consequence, the partition function reads

\begin{align}
Z(\bm{\theta})=\sum_{\substack{\mathbf{B}\in\mathbb{B}\\(|\mathbf B|=|\mathbf B^*|)}}\prod_{i=1}^N\prod_{\alpha=1}^Me^{-(\beta_i+\delta_\alpha)b_{i\alpha}^+}=\prod_{i=1}^N\prod_{\alpha=1}^M\sum_{b_{i\alpha}=-1,1}\left[e^{-(\beta_i+\delta_\alpha)b_{i\alpha}^+}\right]^{|b_{i\alpha}^*|}=\prod_{i=1}^N\prod_{\alpha=1}^M\left[1+e^{-(\beta_i+\delta_\alpha)}\right]^{|b_{i\alpha}^*|}
\end{align}
and leads to the expression

\begin{equation*}
P_\text{BiSCM-FT}(\mathbf{B})=\prod_{i=1}^N\prod_{\alpha=1}^M\left[(1-p_{i\alpha}^+)^{1-b_{i\alpha}^+}(p_{i\alpha}^+)^{b_{i\alpha}^+}\right]^{|b_{i\alpha}^*|},
\end{equation*}
having posed

\begin{equation*}
p_{i\alpha}^+=\frac{e^{-(\beta_i+\delta_\alpha)}}{1+e^{-(\beta_i+\delta_\alpha)}}=\frac{x_iz_\alpha}{1+x_iz_\alpha},
\end{equation*}
where $x_i=e^{-\beta_i}$ and $z_{\alpha}=e^{-\delta_\alpha}$ (the probability that the connected nodes $i$ and $\alpha$ are linked by a negative edge is $p_{i\alpha}^-=1-p_{i\alpha}^+$). Hence, according to the BiSCM-FT, each entry of a signed network such that $|b_{i\alpha}^*|=1$ is a random variable following a Bernoulli distribution that obeys the finite scheme

\begin{align}
b_{i\alpha}\sim
\begin{pmatrix}
-1 & +1\\
p_{i\alpha}^- & p_{i\alpha}^+
\end{pmatrix},\quad\forall\:i,\alpha\text{\:\:s.t.\:\:}|b_{i\alpha}^*|=1.
\end{align}

To determine the parameters that define the BiSCM-FT, let us maximise the likelihood function

\begin{equation}
\mathcal{L}_\text{BiSCM-FT}(\mathbf{x},\mathbf{z})=\ln P_\text{BiSCM-FT}(\mathbf{x},\mathbf{z})=\sum_{i=1}^Nk_i^+(\mathbf{B}^*)\ln(x_i)+\sum_{\alpha=1}^Mh_\alpha^+(\mathbf{B}^*)\ln(z_\alpha)- \sum_{i=1}^N\sum_{\alpha=1}^M|b_{i\alpha}^*|\ln(1+x_iz_\alpha)
\end{equation}
with respect to $x_i$ and $z_\alpha$. Upon doing so, we obtain the system of equations

\begin{align*}
\frac{\partial\mathcal{L}_\text{BiSCM-FT}(\{x_i\}_{i=1}^N,\{z_\alpha\}_{\alpha=1}^M)}{\partial x_i}&=\frac{k_i^+(\mathbf{B}^*)}{x_i}-\sum_{\alpha=1}^M|b_{i\alpha}^*|\frac{z_\alpha}{1+x_iz_\alpha},\quad\forall\:i,\\
\frac{\partial\mathcal{L}_\text{BiSCM-FT}(\{x_i\}_{i=1}^N,\{z_\alpha\}_{\alpha=1}^M)}{\partial z_\alpha}&=\frac{h_\alpha^+(\mathbf{B}^*)}{z_\alpha}-\sum_{i=1}^N|b_{i\alpha}^*|\frac{x_i}{1+x_iz_\alpha},\quad\forall\:\alpha.
\end{align*}
Equating them to zero leads us to find

\begin{align*}
k_i^+(\mathbf{B}^*)&=\sum_{\alpha=1}^M|b_{i\alpha}^*|\frac{x_iz_\alpha}{1+x_iz_\alpha}=\sum_{\alpha=1}^M|b_{i\alpha}^*|p_{i\alpha}^+=\langle k_i^+\rangle,\quad\forall\:i,\\
h_\alpha^+(\mathbf{B}^*)&=\sum_{i=1}^N|b_{i\alpha}^*|\frac{x_iz_\alpha}{1+x_iz_\alpha}=\sum_{i=1}^N|b_{i\alpha}^*|p_{i\alpha}^+=\langle h_\alpha^+\rangle,\quad\forall\:\alpha.
\end{align*}

\clearpage

\section{Numerical optimization of the likelihood function}\label{AppD}

In order to numerically solve the systems of equations defining the BiSCM and the BiSCM-FT, we can follow the guidelines provided in~\cite{vallarano2021}: more specifically, we will adapt the iterative recipe provided there to our (bipartite signed) setting. First, let us notice that the system of equations defining the BiSCM can be re-written as

\begin{align}
x_i&=\frac{k_i^+(\mathbf B^*)}{\sum_{\alpha=1}^M\frac{z_\alpha}{1+x_iz_\alpha+y_iw_\alpha}}\:\Longrightarrow\: x_i^{(n)}=\frac{k_i^+(\mathbf B^*)}{\sum_{\alpha=1}^M\frac{z_\alpha^{(n-1)}}{1+x_i^{(n-1)}z_\alpha^{(n-1)}+y_i^{(n-1)}w_\alpha^{(n-1)}}},\quad\forall\:i,\\
y_i&=\frac{k_i^-(\mathbf B^*)}{\sum_{\alpha=1}^M\frac{w_\alpha}{1+x_iz_\alpha+y_iw_\alpha}}\:\Longrightarrow\: y_i^{(n)}=\frac{k_i^-(\mathbf B^*)}{\sum_{\alpha=1}^M\frac{z_\alpha^{(n-1)}}{1+x_i^{(n-1)}z_\alpha^{(n-1)}+y_i^{(n-1)}w_\alpha^{(n-1)}}},\quad\forall\:i,\\
z_\alpha&=\frac{h_\alpha^+(\mathbf B^*)}{\sum_{i=1}^N\frac{x_i}{1+x_iz_\alpha+y_iw_\alpha}}\:\Longrightarrow\: z_\alpha^{(n)}=\frac{h_\alpha^+(\mathbf B^*)}{\sum_{i=1}^N\frac{x_i^{(n-1)}}{1+x_i^{(n-1)}z_\alpha^{(n-1)}+y_i^{(n-1)}w_\alpha^{(n-1)}}},\quad\forall\:\alpha,\\
w_\alpha&=\frac{h_\alpha^-(\mathbf B^*)}{\sum_{i=1}^N\frac{y_i}{1+x_iz_\alpha+y_iw_\alpha}}\:\Longrightarrow\: w_\alpha^{(n)}=\frac{h_\alpha^-(\mathbf B^*)}{\sum_{i=1}^N\frac{y_i^{(n-1)}}{1+x_i^{(n-1)}z_\alpha^{(n-1)}+y_i^{(n-1)}w_\alpha^{(n-1)}}},\quad\forall\:\alpha;
\end{align}
analogously, the system of equations defining the BiSCM-FT can be re-written as

\begin{align}
x_i&=\frac{k_i^+(\mathbf B^*)}{\sum_{\alpha=1}^M|b_{i\alpha}^*|\frac{z_\alpha}{1+x_iz_\alpha}}\:\Longrightarrow\: x_i^{(n)}=\frac{k_i^+(\mathbf B^*)}{\sum_{\alpha=1}^M|b_{i\alpha}^*|\frac{z_\alpha^{(n-1)}}{1+x_i^{(n-1)}z_\alpha^{(n-1)}}},\quad\forall\:i,\\
z_\alpha&=\frac{h_\alpha^+(\mathbf B^*)}{\sum_{i=1}^N|b_{i\alpha}^*|\frac{x_i}{1+x_iz_\alpha}}\:\Longrightarrow\: z_\alpha^{(n)}=\frac{h_\alpha^+(\mathbf B^*)}{\sum_{i=1}^N|b_{i\alpha}^*|\frac{x_i^{(n-1)}}{1+x_i^{(n-1)}z_\alpha^{(n-1)}}},\quad\forall\:\alpha.
\end{align}

In order for each iterative recipe to converge, an appropriate vector of initial conditions needs to be chosen: hereby, we have opted for the values $x_i=k_i^+(\mathbf{B}^*)/\sqrt{L^+(\mathbf{B}^*)}$ and $y_i=k_i^-(\mathbf{B}^*)/\sqrt{L^-(\mathbf{B}^*)}$, $\forall\:i$; $z_\alpha=h_\alpha^+(\mathbf{B}^*)/\sqrt{L^+(\mathbf{B}^*)}$ and $w_\alpha=h_\alpha^-(\mathbf{B}^*)/\sqrt{L^-(\mathbf{B}^*)}$, $\forall\:\alpha$. Besides, we have adopted two different stopping criteria: the first one is a condition on the Euclidean norm of the vector of differences between the values of the parameters at subsequent iterations, i.e. $||\Delta\vec\theta||_2=\sqrt{\sum_{i=1}^N(\Delta\theta_i)^2}\le10^{-8}$; the second one is a condition on the maximum number of iterations, set to $10^3$.

The accuracy of our method in estimating the constraints has been evaluated by computing the \textit{maximum absolute error} (MAE), defined as the infinite norm of the vector of differences between the empirical and the expected values of the constraints,

\begin{align}
\text{MAE}=\max_{i,\alpha}\left\{|k_i^+(\mathbf B^*)-\langle k_i^+\rangle|,\:|k_i^-(\mathbf B^*)-\langle k_i^-\rangle|,\:|h_\alpha^+(\mathbf B^*)-\langle h_\alpha^+\rangle|,\:|h_\alpha^-(\mathbf B^*)-\langle h_\alpha^-\rangle|\right\}
\end{align}
and the \textit{maximum relative error} (MRE), defined as the infinite norm of the vector of relative differences between the empirical and the expected values of the constraints,

\begin{align}
\text{MRE}=\max_{i,\alpha}\left\{\frac{|k_i^+(\mathbf B^*)-\langle k_i^+\rangle|}{k_i^+(\mathbf B^*)},\:\frac{|k_i^-(\mathbf B^*)\langle k_i^-\rangle|}{k_i^-(\mathbf B^*)},\:\frac{|h_\alpha^+(\mathbf B^*)-\langle h_\alpha^+\rangle|}{h_\alpha^+(\mathbf B^*)},\:\frac{|h_\alpha^-(\mathbf B^*)-\langle h_\alpha^-\rangle|}{h_\alpha^-(\mathbf B^*)}\right\}.
\end{align}

Table~\ref{tab:iterative} shows the time employed by our algorithm to converge as well as its accuracy in reproducing the constraints defining the BiSCM and the BiSCM-FT. Overall, our method is fast and accurate: the numerical errors never exceed $O(10^{-1})$ and the time employed to achieve such an accuracy is always less than a minute. We also provide the basic statistics for the real-world bipartite signed networks considered in the present contribution.

\begin{table}[t!]
\centering
\begin{tabular}{l|c|c|c|c|c|c|c|c|c|c|c|c}
\hline
\multicolumn{7}{c|}{} & \multicolumn{3}{c|}{BiSCM} & \multicolumn{3}{c}{BiSCM-FT} \\
\hline
\hline
& $N$ & $M$ & $L$ & $L^+$ & $L^-$ & $c$ & MAE & MRE & Time (s) & MAE & MRE & Time (s) \\
\hline
\hline
FilmTrust & 1508 & 2071 & 35494 & 28580 & 6914 & $\simeq$ 1.1$\cdot10^{-2}$ & $\simeq 4.7\cdot10^{-1}$ & $\simeq 3.2\cdot10^{-2}$ & $\simeq 1.9$ & $\simeq 2.2\cdot10^{-1}$ & $\simeq 2.3\cdot10^{-2}$ & $\simeq 28$ \\
\hline
U.S. Senate & 145 & 1056 & 27083 & 14979 & 12104 & $\simeq 1.8\cdot10^{-1}$ & $\simeq 2.8\cdot10^{-1}$ & $\simeq 1.5\cdot10^{-2}$ & $\simeq 0.1$ & $\simeq 2.0\cdot10^{-1}$ & $\simeq 2.0\cdot10^{-2}$ & $\simeq 0.1$ \\
\hline
U.S. House & 515 & 1281 & 114378 & 61720 & 52658 & $\simeq 1.7\cdot10^{-1}$ & $\simeq 3.7\cdot10^{-1}$ & $\simeq 1.5\cdot10^{-2}$ & $\simeq 0.5$ & $\simeq 3.4\cdot10^{-1}$ & $\simeq 1.6\cdot10^{-2}$ & $\simeq 0.5$ \\
\hline
\end{tabular}
\caption{\label{tab:iterative} Performance of the fixed-point algorithm to solve the systems of equations defining the BiSCM and the BiSCM-FT on three bipartite real-world networks, i.e. \textit{U.S. Senate}, \textit{U.S. House of Representatives} and \textit{FilmTrust}). $N$ is the number of nodes on the first layer, $M$ is the number of nodes on the second layer, $L$, $L^+$, $L^-$ are the overall number of links, positive links, negative links and $c=L/(N\cdot M)$ is the connectance.}
\end{table}

\clearpage

\section{Comparing partitions}\label{AppE}

Let us, first, distinguish between \textit{Partition I} and \textit{Partition II}: then, TP counts the number of true positives, i.e. the pairs of nodes belonging to the same cluster in both partitions, TN counts the number of true negatives, i.e. the pairs of nodes separated in both partitions, FP counts the number of false positives, i.e. the pairs of nodes belonging to the same cluster in \textit{Partition II} but not in \textit{Partition I} and FN counts the number of false negatives, i.e. the pairs of nodes belonging to the same cluster in \textit{Partition I} but not in \textit{Partition II}.

These concepts lead us to define the following three indices. The \textit{Wallace index} (WI) reads

\begin{align}
\text{WI}&=\frac{\text{TP}}{\text{TP}+\text{FP}}
\end{align}
and measures the agreement between two partitions by quantifying the probability that a pair of nodes in the same block according to \textit{Partition II} is also in the same block according to \textit{Partition I}. The \textit{Rand Index} (RI) reads

\begin{align}
\text{RI}&=\frac{\text{TP}+\text{TN}}{\text{TP}+\text{FP}+\text{TN}+\text{FN}}
\end{align}
and measures the agreement between two partitions by quantifying the percentage of pairs of nodes that are clustered and separated according to both partitions. The \textit{Jaccard Index} (JI) reads

\begin{align}
\text{JI}&=\frac{\text{TP}}{\text{TP}+\text{FP}+\text{FN}}
\end{align}
and measures the agreement between two partitions according to a criterion that is similar to the one informing the WI, although being more severe towards misplaced pairs of nodes. Each index ranges between 0 and 1, with 0 indicating maximum difference and 1 indicating perfect similarity.

\begin{table}[t!]
\centering
\begin{tabular}{l|l|c|c|c}
\hline
Dataset & Projection & Wallace Index & Rand Index & Jaccard Index \\
\hline
\hline
\multirow{3}{*}{U.S. Senate}
& Zero-deflated na\"ive & 0.8049 & 0.6580 & 0.3046 \\
& Zero-deflated global filter & 0.8356 & 0.6041 & 0.2060 \\
& Zero-deflated local filter & 0.5460 & 0.6462 & 0.2125 \\
\hline
\hline
\multirow{3}{*}{U.S. Senate}
& Zero-inflated na\"ive & 0.9515 & 0.4666 & 0.4121 \\
& Zero-inflated global filter & 0.5103 & 0.5065 & 0.2870 \\
& Zero-inflated local filter & 0.4106 & 0.6024 & 0.2662 \\
\hline
\hline
\multirow{3}{*}{U.S. House of Representatives}
& Zero-deflated na\"ive & 0.5448 & 0.5285 & 0.3673 \\
& Zero-deflated global filter & 0.5349 & 0.5082 & 0.1897 \\
& Zero-deflated local filter & 0.4443 & 0.5179 & 0.1590 \\
\hline
\hline
\multirow{3}{*}{U.S. House of Representatives}
& Zero-inflated na\"ive & 0.9904 & 0.8858 & 0.8767 \\
& Zero-inflated global filter & 0.8216 & 0.6887 & 0.4948 \\
& Zero-inflated local filter & 0.6827 & 0.6663 & 0.2104 \\
\hline
\end{tabular}
\caption{Comparison between the partitions found by minimising BIC and by minimising $F$ on our projections of the \textit{U.S. Senate} and \textit{U.S. House of Representatives} datasets.}
\label{tab:performance_indices}
\end{table}

Let us, now, compare $C_\text{BIC}$, considered as the ground truth partition, with $C_\text{F}$, considered as the alternative partition. What we found is that $\text{WI}>\text{RI}>\text{JI}$, a result suggesting that the two recipes for partitioning projections agree on the nodes to be clustered together while disagree on those to be separated. More specifically, \textit{i)} the larger the number of positive links, the better the alignment of the two algorithms for what concerns the pairs of nodes to be clustered together - in fact, WI is closer to 1 for the projections populated by a larger number of positive links; \textit{ii)} the larger the number of positive links, the worse the alignment of the two algorithms for what concerns the pairs of nodes to be separated - in fact, clustering nodes linked by positive ties may cause TN to vanish and FN to become large which, in turn, lead to $\text{WI}\gg\text{RI}=\text{JI}$; \textit{iii)} the larger the number of negative links, the worse the alignment of the two algorithms for what concerns the pairs of nodes to be clustered together - in fact, WI is smaller than RI (becoming `driven' by TN) for the projections induced by local statistical benchmarks, known to enhance the presence of negative ties. Additional evidence is provided by the fact that JI is sistematically smaller than WI, even if they differ for just one term: since, however, the latter is precisely FN, minimising $F$ leads to a large number of `false negatives' - likely caused by the singletons (see also Table~\ref{tab:performance_indices}).

\end{document}